\documentclass[reqno]{article}
\newcommand{\ds}{\displaystyle}
\newcommand{\dsf}{\ds\frac}

\newcommand{\beq}{\begin{equation}}
\newcommand{\eeq}{\end{equation}}

\usepackage{ae} 
\usepackage[T1]{fontenc}
\usepackage[ansinew]{inputenc}
\usepackage{amsmath}
\usepackage{graphicx}
\usepackage{color}
\usepackage[colorlinks]{hyperref}
\usepackage{multicol}
\begin{document}
\small

\begin{center}
{\bf Magnetothermal instabilities in type II superconductors}\
\vskip 0.5cm
{N.A.\,Taylanov}\\
\vskip 0.5cm {\em \footnotesize National University of Uzbekistan}
\end{center}
\begin{center}
\vskip 0.5cm {\bf Abstract} \vskip 0.5cm
\end{center}
\begin{center}
\mbox{\parbox{14cm}{\footnotesize   Magnetothermal instabilities
are one of the peculiar phenomena of interest in conventional
type-II, as well as in high-$T_c$ superconductors. In the present
paper we attempt to analyze the nature and origin of the
magnetothermal instabilities of the critical state and flux jumps
phenomena in superconductors in the light of recent theoretical
and experimental results.}}
\end{center}


\begin{multicols}{2}{

\vskip 0.5cm
\begin{center}
{\bf  Introduction}
\end{center}

    The magnetic flux penetration into a type-II superconductor
occurs in the form of quantized vortices. Each of these vortices
has a normal core, which presents a long cylinder with a radius
comparable to the superconducting coherence length $\xi$.
Undamping superconducting current flows around this cylinder. The
vortex current seizes an area of radius of order of penetration
depth $\lambda$. Each vortex carries one magnetic flux quantum,
which determined by the equality $\phi_0=2\div 10^7 \quad Gs\cdot
sm^2$. The penetration of the vortex into the superconductor
becomes energetically favorable at the field $H_{c1}$. The
vortices are located one after another on the distance $\lambda$
forming correct triangular lattice in the superconductor at the
interval $H_{c1}\ll H\ll H_{c2}$; where $H_{c1}$ and $H_{c2}$ are
the lower and upper critical magnetic fields. With increasing
magnetic field the period of lattices decreases and the vortex
density increases. At the field $H_{c_2}$ the vortex density
becomes so great that the period of lattices becomes an order of
coherence length $\xi$. This means that magnetic flux completely
penetrates into the superconductor and second type phase
transition to the normal state occurs. In the interval of fields
$H_{c1}\ll H\ll H_{c2}$ superconductor is in the mixed state. In
the presence of different types of defects or pinning centers in
the superconductor sample, the vortices may be attached to such
defects. A nature of interaction between the vortices and the
structural defects is determined by the pinning force $F_P$. If
transport current with the density $j$ is passed through
superconductor, the interaction of the current with vortex lines
leads to the emergence of the Lorentz force $F_L$, acting on each
one of the vortices. Under the effect of the Lorentz force $F_L $
the viscous flux flow of vortices begin to move. The viscous
magnetic flux flow in accordance with electromagnetic induction
creates a vortex electric field $E$. This means that energy
dissipation occurs, an electric resistance appears and the
superconductor undergoes a transition to the resistive or to the
normal state. Propagating magnetic flux causes Joule heating,
giving rise to global and/or micro flux avalanches in the critical
state of type-II superconductors. Thus flux jumps results in a
large-scale flux avalanches in a superconductor and their origin
are related to the magnetothermal instabilities.

Magnetothermal instabilities of the critical state and flux jump
phenomenon in hard superconductors with high values of the
critical current density and the critical magnetic fields have
been investigated since 1960's in a classical works of Bean and et
al. [1-3]. A detailed qualitative analysis of the magnetic
instabilities in type-II superconductors was given by Wipf [4].
The magnetic flux jumps were studied in Refs. [5, 6] for a linear
voltage-current characteristics of superconductor in the framework
of adiabatic approximation i.e., assuming that the thermal
diffusion, is much smaller than the magnetic flux diffusion. The
criterion for the stability of the critical state in the case of
dynamic approximation was obtained by Kremlev [7]. The general
concept of the magnetic instabilities in type-II superconductors
was developed in literature [8]. The thermal and electromagnetic
processes, whose development leads to the flux jump, have been
investigated in detail and the stability criterion have been found
in the framework of adiabatic and dynamic approximations in the
viscous flux flow regime by Mints and Rakhmanov [9]. The detailed
theoretical analyze of the flux jumping in the flux creep regime
where the current-voltage characteristics of type-II
superconductors is a nonlinear have been carried out recently by
Mints [10] and by Mints and Brandt [11].

The dynamics of magnetothermal instabilities has been recently
extensively studied in hard type-II as well as high $T_c$
superconductors. And several contrasting models have been proposed
for the detailed description for the nature of their origin and
dynamics. The dynamics and the nature of observed in recent
magnetization measurements of various techniques flux jumps can be
explained in the scenario of magnetothermal instability theory
[1]. According to this theory moving flux lines increase the local
temperature $T$ and therefore reduce the critical current density
$j_c$, which triggers further flux motion. This positive feedback
may result in a magnetic flux avalanches and thermal instabilities
in the superconductor sample. It is notice able that a recent
theoretical and experimental investigations show that the dynamics
of these flux jump instabilities strongly depends on the local
temperature change, the rate of field change, the critical current
density and its temperature and field derivatives, the ratio
between thermal conductivity and heat capacity, the sample
geometry, cooling conditions, pinning and micro-structural
properties of a considered superconducting material.

Many experimental and theoretical investigations on magnetothermal
instabilities in low and high-$T_c$ superconducting materials with
different shapes and geometries, including thin $Nb, MgB_2$ films,
organic superconductors and layered compounds, single and textured
crystals, rings and bulk samples have been recently reported.
However, no systematic studies of this problem have been reported
and there is little understanding as to how different internal and
external parameters affect flux jumping behaviors in a various
superconducting samples.

Interestingly that most magnetization measurements show that the
flux jumps occur at lower magnitudes of temperature and magnetic
fields in a single superconductor material. Some experimental
results show that flux jumps occur at temperatures and magnetic
fields larger than the full penetration field $H_p$, but lower
than the $H_{c1}$. For instance, magnetothermal instabilities in
organic type-II superconductors have been observed in the form of
abrupt flux jumps at extremely low temperatures by Mola et al.
[12] by means a torque magnetometer. Based on analysis of the
temperature dependence of the jumps they found that the amplitude
of the observed flux jumps increases with temperature as $A\sim
T^{3/2}$, which consistent with accepted models. They have showed
that as the temperature is increased the flux jumps are observed
less frequently, and with greater magnitude. Thus, the authors
believed that the sudden cessation of the flux jumps above a
characteristic field $B_j$, which decreases upon increasing the
temperature. It was also found that the average amplitude of the
flux jumps is an increasing function of angle having an
$A(\Theta)\approx 1-\cosh(\Theta)$ dependence.

Similar temperature and field dependencies of the flux jump
instabilities have been observed by Radovan and Zieve [13] in
their local Hall probe measurements. They found both large and
small flux jumps at different temperatures in a type-II thin Pb
film superconductors. The flux jumps occur only at very low
applied fields. A large jumps observed only at relatively high
temperatures. It has been shown by these authors that at
sufficiently low temperatures the jump size decreases roughly as
$T^3$, approaching a finite value at zero temperature. They found
a power law distributions of avalanche size at temperatures
$T=0.3$ K and $T=4.3$ K with corresponding exponents of 2.01 and
1.09, respectively. Thus the authors argued that an interplay
between vortex density and the microstructure is at the origin of
the flux instabilities.

The strong temperature dependence of the magnitude and frequency
of flux jumps have been observed by Nowak and his co-workers [14]
in a ring-shaped Nb thin films using a Hall probes. The
magnetization of the film was measured at $T=1.4\div 10$ K as a
function of temperature and applied field, perpendicular to the
plane of the sample. At $T\sim 3.1$ K has been detected a
crossover from a broad to narrow of the size distribution of
internal avalanches in the rings. The authors argued that the
small value of specific heat at low temperatures is at the origin
of these jumps in magnetization curves. So, these low-temperature
huge avalanches in the rings may be qualitatively understood in a
thermal instability scenario [9], according which at low
temperatures the vortex motion can result in a local temperature
rise. The temperature rise leads to a decrease in the critical
current density and therefore magnetothermal instability is
produced. The stability parameter $\beta$ [8, 9] numerically has
been calculated by using the measured temperature dependencies of
the critical current density $j_c$ and heat capacity $\nu\sim
T^3$. The authors found that at highest temperatures thermal
effects are negligible and a largest in magnitude heat capacity
serves as a stabilizing factor of the Bean's critical state.

A such type of flux jumps in the magnetization curves at low
temperatures for a single YBaCuO crystals with a different
transition temperatures $T_c$ has been found by Khene and Barbara
[15]. It was shown that the first flux jump field $B_j$ depends
not only on critical current density $j_c$ and specific heat
$\nu$, but also on the field sweep rate and the transition
temperature $T_c$ of the sample. The results of their
magnetization measurements show that the field for the first flux
jump $B_j$ increases as temperature increases and the flux jumps
disappear at higher temperatures. The results obtained in this
work suggest a better thermal and magnetic stability for
considered single crystals with low transition temperature $T_c$.
The effect of magnetic field sweep rate on the behavior of flux
jumps has also been investigated. The field $B_j$ decreases with
increasing magnetic field sweep rate $\dot B_e$. Interestingly
that these results are very similar to those obtained in
conventional II-type superconductors [16].

A more detailed the temperature and magnetic field sweep rate
dependencies of flux jumping has been analyzed by Nabialek et al.
[17] at temperature of 4.2 K in polycrystalline single crystals by
means of magnetization, screening and torque measurements. It has
been shown that as the sweep rate decreases the value of the first
flux jump field $B_j$ increases rapidly and tends to saturate at
higher sweep rates. Similar strong sweep rate dependence on the
first flux jump field $B_j$ early has been observed in other
experiments of McHenry [18] and Guillot [19]. Nabialek et al.
[17], also investigated the temperature dependence of flux
jumping, as well as the influence of flux creep and demagnetizing
effects on the critical state stability.

The behavior of flux jumps with different amplitudes in
superconductors have studied Vanacken et al. [20] using pulsed
field magnetization measurements. The time dependence of the flux
jumps was presented. A periodical jumps were observed for both the
positive and negative field polarities. Surprisingly that the
sharp flux jumps occur at lowering magnitude of the field for all
temperatures below $T=30$ K. These jumps are expected to be
equidistant  $B_j\approx 0.3$ T. As has been shown that the flux
jumps strongly depend on the magnetic field sweep rate and the
magneto-thermal history of the sample, on the temperature and the
critical current density. Such dependence of occurrence of flux
jump field on the sweep rate of external magnetic field and the
temperature for the melt-textured superconductor materials has
been studied by Fuchs et al. [21] at temperatures $T>20$ K. The
observed magnetic instabilities were discussed within the
framework of existing theoretical results.

Recently Zhao et al. [22] have observed many multiple small and
irregular flux jump instabilities in $MgB_2$ thin films at
temperature below 2 K and applied field of 1.3 T by means a SQUID
magnetometer measurements. The magnitude of these instabilities
are much larger in low fields than in high fields. With increasing
of temperature the small instabilities evolved into some larger
ones and disappear completely at temperature $T=14$ K. The authors
believed that there are many places for the avalanche to occur due
to a high density of small defects formed during the preparation
process of the thin films leading to much stronger critical
current densities. As pointed authors thermal diffusion is much
easier in thin films due to their very small thickness and large
surface area exposed to the environment and consequently, in thin
films each avalanche is small in magnitude but the number of
avalanches can be huge. The authors argued that this simple
picture may give an explanation to many small flux jumps observed
in superconducting thin films. They reported the observation of
the suppression of the critical current density at low
temperatures due to many small jumps in thin films.

Muller and Andrikidis [23] using a SQUID magnetometer has measured
the magnetization loop for a melt-textured high $T_c$
superconductors at different temperature and magnetic field
intervals. The first flux jump occur at temperature T=3.0 K in the
magnetization loop, when the applied field is perpendicular to the
crystallographic c axis of the sample. At temperature T=3.5 K
solitary jumps were found and above 3.9 K no flux jumps occur in
the magnetization curve. With increasing temperature the first
jump field increases and at higher temperatures (above T=3.05 K)
flux jump suddenly disappears. The measured values of $B_j$ has
been compared with existing theoretical results [8, 9]. In this
work the stability criterion for flux jumps has been analytically
calculated using Kim-Anderson critical state model.

Chabanenko at al. [24] have studied magnetothermal instabilities
and giant flux jumps, both theoretically and experimentally in the
framework of adiabatic approximation using various dependencies of
the critical current density on the magnetic field. In particular,
they have numerically calculated magnetization and
magnetostriction loops with flux jumps employing the Kim-Anderson
critical state model and exponential model for the dependence
$j_c(H)$. On the basis of these theoretical calculations it has
been constructed a H-T phase diagram of flux jump instabilities.
The authors showed that the maximum value of temperature after the
flux jump strongly depends on the magnetic field sweep rate and
micro-structural properties of the sample.

Vasilev et al. [25] experimentally studied the time of duration of
the flux jumps and amount of the magnetic flux entering into a
disc type of superconductor sample as a function of temperature
and external magnetic field. In another paper Chabanenko et al.
[26] have studied the structure of the flux jumps in different
superconducting samples, such as Nb-Ti samples, polycrystalline Nb
plates and melt-textured YBaCuO slabs on the basis of a temporal
dependence of the surface magnetic induction using a miniature
Hall probe sensors. The authors have observed very interesting
oscillating phenomena originated by the thermomagnetic avalanches
in the vortex matter for both low and high-$T_c$ superconductor
samples. These oscillation can be qualitatively interpreted in
terms of the theoretical model which takes into account the
existence of a definite value of the effective vortex mass, i.e
the inertial properties of the vortex system [27]. In this work a
details of three stages of the thermomagnetic instability
development was also proposed.

The influence of the time-dependent external conditions on the
stability threshold of the critical state for hard type-II
superconductors has been studied theoretically by Mints and
Rakhmanov [28]. They found a stability criterion of the critical
state as a function of the rate of variation of the external
magnetic field and the external temperature in hard
superconductors. The authors argued that the existence of a
background electric field induced by a variable external magnetic
field is essential for observing a temperature and electric field
oscillations in a superconductor sample. Similar magnetothermal
oscillations in the form of flux jumps have been observed by Kumm
et al. [29] in a melt-textured crystals in external varying
magnetic field. The interval $\Delta B_e$ between consecutive
jumps increases with sweep rate $\dot B_e$ and decreases with
temperature. Every jump is preceded by oscillations of the induced
voltage and the sample temperature. A similar oscillations
preceding the flux jump instabilities have been studied by
numerous researchers in earlier both experimentally by Shimamoto
[30] and theoretically by Maksimov and Mints [31]. Earlier Zebouni
et al. [32] have detected that the period of oscillations
increases strongly with increasing temperature. The oscillations
disappear sharply if the sweep was stopped. This effect due to the
simultaneous and sharp drop of the sample temperature to the bath
temperature. Such a series of heating pulses of very large
amplitude appeared at discrete reproducible values of $B$. The
pulses were observed in decreasing field, but at relatively lower
values of $B$. Simultaneously with each of these heat pulses,
sudden penetration of magnetic flux was observed. Therefore the
pulses related to discontinuous steps in the magnetic flux
penetration - flux jumps. A tentative identification of these
fluctuations with one of the collective modes of vibration of the
vortices predicted by de Gennes and Martison [33]. Such limited
flux jump instabilities, initiated by a varying external magnetic
field perturbation has been observed in ceramic high-$T_c$
superconductors by Bodi et al. [34]. An extensive study of the
magnetothermal oscillations and limited flux jumps in a high-$T_c$
superconductor samples was given by Legrand et al. [35]. Later
Legrand et al. [36] have proposed that the magneto-thermal
oscillations in the Bean's critical state may arise as coupled
oscillations of small perturbations of the temperature and the
electric field. They both theoretically and experimentally studied
the dependence of the frequency of the oscillations on the
magnetic sweep rate and the temperature for a granular
superconductors. The experimental results for magnetothermal
oscillations in the temperature $T=3\div 7$ K and magnetic field
sweep rate $\dot B_e=10\div 45$ G/s ranges show that at low values
of the applied field $B_e(t)$ there is slow temperature increase
due to the small vortex avalanches establishing the critical
state. Above certain threshold magnetic field, will appear
temperature oscillations with the period $t$ in the range $t\sim
10\div 70$ s and its amplitude increasing in time. A flux jump
occurs close to the Bean field accompanied by a temperature rise
up to about 12 K with a characteristic time of the order of 1 s.
The frequency of these oscillations is proportional to the
magnetic field sweep rate $\dot B_e$.

Recently, it has been shown [37] that the local Joule heating due
to planar defects or low-angle grain boundaries can cause thermal
instabilities in a high-$T_c$ coated conductors. It was shown that
the thermal instabilities are more pronounced at lower
temperatures and magnetic fields or under poorer gas cooling. The
author have obtained the stability criterion for a planar defect
in a thin film for a power-law voltage-current characteristics of
the sample on the basis of calculation a steady-state heat balance
equation. It has been pointed by Gurevich [37], also that the
highly nonlinear voltage-current characteristics may cause strong
disturbances of the electric field and dissipation with spatial
size much larger than the defect size in the sample. Thus the
resulting local hot spots near defects can trigger thermal
instability in thin films. At sufficiently low electric fields $E$
the voltage-current characteristics of type-II superconductors is
highly nonlinear and differential conductivity $\sigma$ is a
function of the electric field $E$ [38, 39]. The dependence
$\sigma$ of $E$ at small values of the electric field $E$ may
considerably affect the critical state stability and the
occurrence of the flux jumps in the sample [39]. As recently has
been shown by Mints [10] that the nonlinear differential
conductivity $\sigma$ significantly affects the flux jumping
process and therefore the first flux jump field $B_j$ in the flux
creep regime. In particular, the author derived a criterion for
thermomagnetic flux-jump instability for the case of transverse
geometry of thin films taking into account the nonlinear
dependence of the background electric field on sample conductivity
in the framework of Bean's critical state model. It was shown that
the magnetic field sweep rate affects the instability. This effect
is connected with the non-linear electric field dependence of
$j(E)$ in the flux creep regime. The first flux jump field $B_j$
decreases with increasing sweep rate field $\dot B_e$. As flux
creep is relatively strong in high-$T_c$ superconductors, one may
expect it to have a significant influence on the critical state
stability in these materials.

A detailed measurements have been performed by means a torque and
SQUID magnetometer during magnetic field sweep rate by Monier and
Fruchter [40] in a single crystal of the organic superconductor.
They found an expression for the first flux jump field $B_j$ for a
cylinder sample assuming that the temperature in the sample is
uniform. The obtained criterion for flux jumps for a nonlinear
voltage-current characteristic in the dynamic approximation in a
good agreement with a recent results of Mints [10]. It was  shown
that magnetothermal instabilities may be suppressed by an increase
of heat transfer coefficient or the use of thin samples. The
activation energy, the screening current density and its
temperature derivative has been found from the dynamic relaxation
measurements.

In the present paper, we shall investigate the magnetothermal flux
jump instabilities of the critical state in conventional and
high-$T_c$ superconductors. We determine the flux jump field $B_j$
and critical state stability criterion within the framework of
Bean's model. To determine the flux jump field an analytical
simulations of coupled equations for the magnetic $\vec B (\vec r,
t)$ and electric $\vec E(\vec r, t)$ field inductions and
temperature $T(\vec r, t)$ has been performed. The field of the
first flux jumps $B_j$ is calculated analytically by using the
dynamic and the adiabatic approximations. The qualitative analysis
of the magnetic flux jumps instabilities for superconductors will
be provided.

\vskip 0.5cm
\begin{center}
{\bf\S 1. Flux jump instabilities }
\end{center}

Let us suppose that at the initial moment of time, the magnetic
field in the sample is uniform and equal to $B_0$, after which the
external field rises to some value $B_e$. With an increase of the
external field $B_e$, the vortices penetrates the sample. The
viscous flow of vortices inside the sample causes an electric
field $E$ which generates persistent currents near the sample
surface, having a density $j\simeq j_c$ and power dissipation.
This dissipation heats the sample which leads to further changes
in the magnetic flux density and further increase the sample
temperature locally decrease in the critical current density
$j_c$. With a decrease $j_c$, the vortex lines penetrates the
sample more deeply, more heat is released and new dissipation
occurs thus resulting in a new increase of temperature $T$. Under
certain conditions such a dissipative flux motion can cause
magnetothermal instabilities.

Mathematical problem of theoretical study the dynamics of thermal
and electromagnetic perturbations in a superconductor sample in
the flux creep regime can be formulated on the basis of a system
nonlinear diffusion-like equations for the thermal and
electromagnetic field perturbations with account nonlinear
relationship between the field and current in superconductor
sample. The distribution of the magnetic flux density $\vec B$ and
the transport current density $\vec j$ inside a superconductor is
described by the equation

\begin{equation}
rot\vec B=\dsf{4\pi}{c}\vec j.
\end{equation}
When the penetrated magnetic flux changes with time, an electric
field $\vec E$ is generated inside the sample according to
Faraday's law
\begin{equation}
rot\vec E=\dsf{1}{c}\dsf{d\vec B}{dt}.
\end{equation}
The temperature distribution in superconductor is governed by the
heat conduction diffusion equation
\begin{equation}
\nu (T)\dsf{dT}{dt}=\nabla[\kappa(T)\nabla T]+\vec j\vec E,
\end{equation}
Here $\nu=\nu(T)$ and $\kappa=\kappa(T)$ are the specific heat and
thermal conductivity, respectively. The above equations should be
supplemented by a current-voltage characteristics of
superconductors, which has the form
\begin{equation}
\vec j=\vec j_{c}(T,\vec B)+\vec j(\vec E).
\end{equation}
In general, the critical current density depends, on both the
local magnetic field $B$ and local temperature $T$, thus, the
magnetic flux profile is determined by equation $j_c=j_c(B, T)$.
C.P. Bean [2] has proposed the critical state model which
successfully used to describe magnetic and transport properties of
hard superconductors. According to this model the current density
$\vec j$ is equal to the critical current density $\vec j_c$ and
independent of magnetic field. In order to obtain analytical
results of a set Eqs. (1)-(4), we suggest that $j_c$ is
independent on magnetic field $B$ and use the Bean critical state
model $j_c=j_c(B_e, T)$. We suppose that the critical current
density $j_c$ is linearly dependent on the local temperature

\begin{equation}
j_c(T)=j_0-a(T-T_0).
\end{equation}
where $a=j_0/(T_c-T_0)$; $j_0$ is the equilibrium current density;
$T_0$ and $T_c$ are the equilibrium and critical temperatures of
the sample, respectively. The characteristic field dependence of
$j(E)$ in the region of sufficiently strong electric field $(E\geq
E_f)$ can be approximated by the piece-wise linear function
$j=\sigma_f E$; $\sigma_f=\eta c^2/B\Phi_0=\sigma_n H_{c_2}/B$ is
the effective conductivity in the flux flow regime; $\eta$ is the
viscous coefficient, $\Phi_0=\pi h c/2e$ is the magnetic flux
quantum, $\sigma_n$ is the differential conductivity in the normal
state, $H_{c2}$ is the upper critical magnetic field, $E_f$ is the
boundary of the linear area in the voltage-current characteristics
of the sample [41]. Notice, that in the flux flow regime the
differential conductivity $\sigma_f$ is approximately constant,
i.e., independent on the electric field E.
\begin{center}
\includegraphics[width=4.5583in]{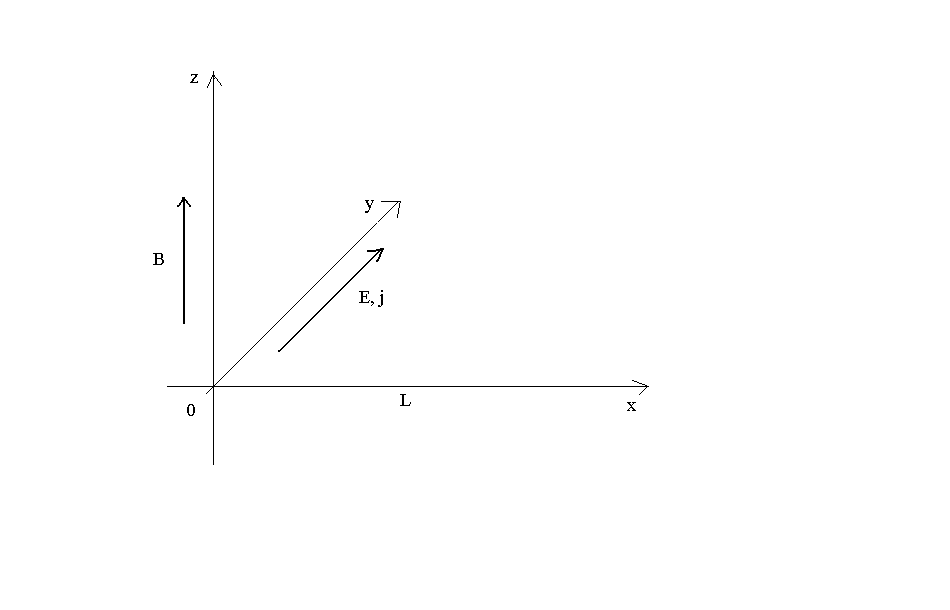}
\end{center}
\begin{center}
Fig.1  The geometry of the problem
\end{center}
Let us formulate a differential equations governing the dynamics
of small temperature and electromagnetic field perturbation in a
superconductor sample. We study the evolution of thermal and
electromagnetic penetration process in a simple geometry -
superconducting semi-infinitive sample $x\geq 0$ (Fig 1.). We
assume that the external magnetic field induction $B_e$ is
parallel to the z-axis and the magnetic field sweep rate
$\dot{B_e}$ is constant. When the magnetic field with the flux
density $B_e$ is applied in the direction of the z-axis, the
transport current $j(x, t)$ and the electric field $E(x, t)$ are
induced inside the slab along the y-axis. For this geometry the
spatial and temporal evolution of small thermal T(x, t) and
electromagnetic field E(x, t) perturbations are described by the
thermal diffusion equation coupled to Maxwell's equations

\begin{equation}
\nu\dsf{dT}{dt}=k\dsf{d^2T}{dx^2}+jE,
\end{equation}
\begin{equation}
\dsf{d^2E}{dx^2}=\dsf{4\pi}{c^2}\left[\dsf{dj}{dE}\dsf{dE}{dt}-\dsf{dj_c}{dT}\dsf{dT}{dt}\right].
\end{equation}
It should be noted that the nonlinear diffusion-type equations (6)
and (7), totally determine the problem of the space-time
distribution of the temperature and electromagnetic field profiles
in the flux flow regime with a linear current-voltage
characteristics in the semi-infinite sample.

The system of differential equations (6) and (7) may be integrated
analytically subject to appropriate initial and boundary
conditions for the thermal and electromagnetic small
perturbations. We present the thermal boundary conditions as
\begin{equation}
\kappa\dsf{dT(0, t)}{dx}=-w(T-T_0),\quad T(L, t)=T_0,
\end{equation}
where w is the heat transport coefficient. Let us assume that the
magnetic field perturbation is equal to zero at the sample surface
and according to relation (2), we obtain the first electrodynamic
boundary condition
\begin{equation}
\dsf{dE(0,t)}{dx}=0.
\end{equation}
The second boundary condition for the electric field E(x, t) at
the flux front $x=L$ can be presented as
\begin{equation}
E(L,t)=0,
\end{equation}
The boundary conditions for the magnetic field induction are
$$
dB(0, t)=B_e,\quad B(L, t)=0,
$$
where $L=\dsf{cB_e}{4\pi j_c}$  is the London penetration depth.

\vskip 0.5cm
\begin{center}
{\bf\S 1.1. Dynamic instability}
\end{center}

In this section we study the dynamics of flux-jump instability of
during flux penetration into superconductor sample within the
dynamical approximation [8]. As we know [8, 9] that a nature of
the flux jumps depends on the competition between diffusive and
dissipative processes through the dimensionless parameter
$$
\tau=\dsf{4\pi\sigma_f\kappa}{c^2\nu}=\dsf{D_t}{D_m}.
$$
where $D_t=\kappa/\nu$ is the thermal diffusivity and
$D_m=c^2/4\pi\sigma_f$ the magnetic diffusivity coefficients,
respectively. Therefore the flux instability criterion is
determined mainly by the relation of the magnetic $D_m$ and
thermal $D_t$ diffusion coefficients. Let us consider the case for
the flux jumps corresponding to the limiting case $\tau\gg 1$.
Consequently, it can be assumed that the initial rapid heating
stage of a flux jump takes place on the background of a
"frozen-in" magnetic flux.  Therefore, under this dynamic
approximation, we obtain from (7) the relation between electric
field $E(x, t)$ and temperature $T(x, t)$ perturbations in the
following form
\begin{equation}
\dsf{dj}{dE}E-\dsf{j_c}{T_c-T_0}T=0
\end{equation}
We notice that the last relation between $E(x, t)$ and temperature
$T(x, t)$ has been derived in the assumption that the decrease of
the critical current density $j_c$ resulting from a temperature
perturbation $T(x, t)$ compensates with increase of the resistive
current density $j$ resulting from an electric field perturbation
$E(x, t)$, so a total current density remains constant [8]. Upon
substituting the expression (11) into the equation (6) and
excluding the variable $E(x, t)$ one can get the differential
equation for the distribution of thermal perturbation, which can
be conveniently presented in the following dimensionless form

\begin{equation}
\dsf{d^2\Theta}{dz^2}+\dsf{d\Theta}{d\tau}+\delta\Theta=0.
\end{equation}
Here we introduced the following dimensionless variables

$$
\Theta=\dsf{T}{T_c-T_0},\quad \tau=\dsf{t}{t_k},\quad t_k=\dsf{\nu
L^2}{\kappa},\quad z=\dsf{x}{L},
$$
and parameters
$$
\quad E_e=\dsf{aL^2}{\kappa},\quad \delta=\dsf{j_c}{\sigma_fE_e}.
$$
The equation (12) can be easily solved by using a method of
separation of variables. Taking into account the thermal boundary
conditions
\begin{equation}
\Theta(1, \tau)=1, \qquad \dsf{d\Theta(0, \tau)}{dz}=0.
\end{equation}
an explicit solution to equation (12) can be written as the
following form

\begin{equation}
T(x,
t)=(T_c-T_0)\exp\left[\left(\delta-\dsf{\pi^2}{4L^2}\right)(t-t_0)\right]\cos\dsf{\pi}{2L}x+T_0
\end{equation}
where $t_0$ is a constant parameter. The distribution of the
electric field perturbation $E(x,t)$ can be found by using
relations (11) and (14)

\begin{equation}
E(x,t)=E_c\exp{\left(\left[\delta-\dsf{
\pi^2}{4L^2}\right](t-t_0)\right)\cosh\dsf{\pi}{2L}x}
\end{equation}
Using the obtained solutions (14) and (15) one can easily find the
critical field $B_j$ starting from which flux jump takes place

\begin{equation}
B_j=\dsf{\pi^2}{2}\left[16\dsf{\kappa}{a}\dsf{\sigma_fj_c}{c^2}\right]^{1/2}
\end{equation}
and the critical thickness $d_c$ of the sample

\begin{equation}
d_c=\dsf{\pi}{2}\sqrt{\dsf{\sigma_f\kappa}{aj_c}}.
\end{equation}
We can easily estimate $B_j$ and $d_c$ using by a typical
numerical values of a well known parameters of the sample which
has been made in many publications (see, for example [8]).

The dynamic approximation which we used here is justified if the
thermal flux diffusion is much greater than the magnetic diffusion
$D_t\gg D_m$. The thermal $D_t=10^3\div 10^4 \cdot cm^2/s$ and
magnetic  $D_m=1\div 10\cdot cm^2/s$ diffusion coefficients we
estimated using by a numerical values of parameters. On the other
hand, using the obtained numerical values of $D_t$ and $D_m$ it is
easily estimate the thermal and magnetic diffusion times, taking
into account that, both $t_{\kappa}$ and $t_m$ are inversely
proportional to the corresponding thermal and magnetic diffusion
coefficients, respectively. Then one can be seen, within the time
interval $t_{\kappa}\ll \Delta t\ll t_m$ the results obtained
provide for a highly accurate description of the dynamical
evolution of temperature and electromagnetic perturbations in the
superconductor sample.

\begin{center}
{\bf\S 1.2. Adiabatic instability}
\end{center}

In the adiabatic approximation we assume that rapid propagation of
the flux line is accompanied by an adiabatic heating of the
superconductor [8]. In this case the thermal conditions have a
little effect on the occurrence of the flux jumps. Therefore
thermal conductivity is negligible and corresponding term can be
neglected in the heat diffusion equation. In this limiting case
$\tau\ll 1$ the distribution of the electric field is described by
the following nonlinear equation

\begin{equation}
 e\dsf{d^3 e}{d\tau dz^2}-\dsf{d^2 e}{dz^2} \dsf{d e}{d\tau}
+\left[\left(\dsf{d e}{d\tau}\right)^2-e
\dsf{d^2e}{d\tau^2}\right]+e^2 \dsf{d^2e}{dz^2}=0.
\end{equation}
$$
\tau=\dsf{t}{t_m},\quad  t_m=\dsf{L^2}{D_m}, \quad
z=\dsf{x}{L},\quad e=\dsf{E}{E_e},\quad E_e=\dsf{\sigma_f\nu}{a}.
$$
The solution of (18) together with the boundary conditions (8) can
be easily obtained analogously, as in a previous section, so we
have

\begin{equation}
e(z, \tau)=\lambda(\tau)\psi(z).
\end{equation}
after substituting the last relation into (18) one obtains the
following expressions for the functions $\phi(x)$ and
$\lambda(\tau)$
\begin{equation}
\dsf{d^2\phi}{dx^2}+k=0,
\end{equation}
\begin{equation}
f\dsf{d^2f}{d\tau^2}-\left(\dsf{df}{d\tau}\right)^2+kf^2=0,
\end{equation}
where k is the constant parameter to be determined. The equation
(20) has an explicit solution
\begin{equation}
\phi(x)=\dsf{k}{2}(1-z^2).
\end{equation}
Let us rewrite the equation (21) in the form of
\begin{equation}
\left(\dsf{df}{d\tau}\right)^2=2f^2(b-kf).
\end{equation}
which has an exact solution in the form
\begin{equation}
f(\tau)=\dsf{b^2}{k} ch^{-2}\dsf{b(\tau-\tau_p)}{\sqrt{2}}.
\end{equation}
where $\tau_p$ is the integrating constant and b is the free
parameter. Combining above solutions (22) and (24) we get the
following expression for the electric field and temperature
distribution

\begin{equation}
E(x,
t)=\dsf{E_e}{2}\dsf{(L^2-x^2)}{ch^{2}\dsf{b(t-t_p)}{\sqrt{2}}}
\end{equation}
\begin{equation}
T(x,t)=\dsf{\sigma_fE_e}{2a}
(L^2-x^2)\left[ch^{-2}\dsf{b(t-t_p)}{\sqrt{2}}-1\right]+
\sqrt{2}E_e\dsf{b}{a}th\dsf{b(t-t_p)}{\sqrt{2}}+T_0.
\end{equation}
According to (25) perturbation $E(x, t)$ decreases exponentially
with time $\Delta t=\sqrt{2}/b$. Motion of flux lines caused by
$E(x, t)$ perturbations leads to increase temperature in the
region $x\leq L$ of the sample. From (26) at $t-t_p\gg\Delta t$ we
obtain the following expression for the maximum heating during the
flux jump
\begin{equation}
\Delta\Theta_m =\dsf{T-T_0}{T_c-T_0}=2\sqrt{\dsf{\sigma_f E_e}{j_c
\beta}}.
\end{equation}
Numerical estimation gives  $\Delta\Theta_m\sim 2\div 3$ for the
typical values of parameters of the sample. Such amount of heating
will be dissipated during the adiabatic flux jump instability
which is determined by a well known stability parameter
\begin{equation}
\beta=\dsf{4\pi L^2j_{c}^{2}}{c^2\nu (T_c-T_0)}.
\end{equation}
A more detailed derivation of the adiabatic stability criterion
for flux jumps within the framework of Bean critical state model
has been given in [8]. To compare the theoretical value of $\beta$
with experimental data, one must know the temperature dependence
the heat capacity coefficient of the sample, which at low
temperatures can be expressed as $\nu(T)\sim \nu_0(T/T_0)^n$,
where $\nu_0$ is a numerical factor, $n\simeq 3.1$. Using the
expression for $\nu(T)$ one can estimate the value of $\beta$
numerically for an existing of various functional dependencies of
$j_c(T)$. It should be noted that due to high values of critical
temperature and thus heat capacity in high $T_c$ superconductors,
the experimental value of the first flux jump field $\beta$ can be
larger than in conventional type II superconductors.

\vskip 4cm
\begin{center}
{\bf\S 1.3. Flux creep}
\end{center}

In general, the critical state stability threshold is determined
by many external and internal factors, as by the sweep rate of the
external magnetic field, the type of voltage-current
characteristics, the critical current density and its magnetic
field and temperature derivatives, the profile temperature and
surface cooling conditions, the sample geometry and pinning
properties of the considered sample. The obtained above results
are valid in the flux flow regime, where voltage current-current
characteristics of hard superconductor is described by linear
dependence of $j(E)$ at sufficiently large values of electric
field [41]. As can be seen from the obtained results for the flux
flow state the value of $B_j$ depends mainly on the critical
current density and the specific heat of the sample. The nonlinear
part of the curve $j(E)$ in the region of weak electric fields is
associated with flux creep. In the flux creep regime the magnitude
of flux jump field $B_j$ strongly depends on variation of external
parameters, in particular, on the magnetic field sweep rate [10,
11]. It can be shown [10, 11] that flux creep may sufficiently
affect on the occurrence of the flux jump instabilities. In this
section we shall discuss the effect of flux creep and the
nonlinear current-voltage characteristics on the threshold of flux
jumps, qualitatively. The current-voltage characteristics of type-
II conventional as well as high-$T_c$ superconductors in the flux
creep regime is a highly nonlinear due to thermally activated
magnetic flux motion. Thermally activated flux motion or flux
creep problem in superconductor samples with various geometries
and conditions has been recently extensively studied by many
researchers [42-64]. According to Kim-Anderson theory [55, 56] the
thermally activated flux motion is described by a well known
expression
\begin{equation}
v=v_0 \exp[-U/kT],
\end{equation}
Here $v_0$ is the velocity of the thermally activated flux motion
at zero temperature, $U$ is the activation energy due to vortex
pinning. The activation energy $U=U(j, B, T)$ depends on
temperature $T$, magnetic field induction $B$ and current density
$j$. The dependence of $U$ on the current density $j$ is
extensively discussed in the literature [57]. In particular,
Burlachkov et al. [58] have analyzed the flux creep at different
dependencies of the activation energy $U$ on field $B$ and current
density $j$. For the vortex glass and collective creep models the
potential barriers highly nonlinear function of $j$ [59]. A
voltage-current characteristic of a type-II superconductor in the
flux creep state is characterized by the power or exponential law
increase of $E$ with increasing $j$. For the linear current
dependence of the potential barrier $U(j)$ [55, 56], the
dependence $j(E)$ has the form
\begin{equation}
\vec j=j_c+j_1\ln\left[\dsf{\vec E}{E_c}\right],
\end{equation}
where parameter $j_1$ determines the slope of the j-E curve and it
is assumed $j_1\ll j_c$. In this case the differential
conductivity $\sigma$  is determined by the following expression
(see, Refs. [10, 11])
\begin{equation}
\sigma=\dsf{dj}{dE}=\dsf{j_1}{E}.
\end{equation}
For the logarithmic current dependence of the potential barrier
$U(j)$ proposed by Zeldov et.all. [59]
\begin{equation}
U(j)=U_0 \ln\left[\dsf{j_{c}}{j}\right]^n,
\end{equation}
the dependence $j(E)$ has the form
\begin{equation}
\vec j=j_c\left[\dsf{\vec E}{E_c}\right]^{1/n},
\end{equation}
when the flux creep is determined by numerous spatial defects of
the sample. $U_0$=const and $E_c$ is the crossover electric field.
Here the parameter $n=U_0/kT$ is a function of temperature $T$,
magnetic field $H$ and depends on the pinning regimes and can vary
widely for various types of superconductors. In the case $n=1$ the
power-law relation (33) reduces to Ohm's law, describing the
normal or flux-flow regime. For infinitely large $n$, the equation
describes the Bean critical state model $j=j_c$ [1]. When
$1<n<\infty$, this equation describes nonlinear flux creep [59].
In this case the differential conductivity $\sigma$ is determined
by the following expression

\begin{equation}
\sigma=\dsf{dj}{dE}=\dsf{j_c}{nE}.
\end{equation}
It is assumed, for simplicity, that the value of n temperature and
magnetic-field independent. It should be noted that the nonlinear
diffusion-type equations (6) and (7), completed by the flux creep
equation (34), totally determine the problem of the space-time
distribution of the temperature and electromagnetic field profiles
in the flux creep regime with a nonlinear current-voltage
characteristics in a semi-infinite superconductor sample.

It should be noted that the investigation of the stability
conditions in the flux creep regime is very difficult due to
absence of solution of system equations (6)-(7) together with
nonlinear $j(E)$ dependence. However, in some limiting cases, it
can be solved the problem if we take into account that the heating
due to viscous flux motion is negligibly small and get some
approximate solution, describing the evolution of thermal and
magnetic field diffusion in the creep regime.  According to
relation (34) the conductivity decreases with the increasing of
electric field $E$, while $\sigma\sim E^{-1}$ and it strongly
depends on the external magnetic field sweep rate $E\sim \dot
B_e$. Therefore the stability criterion also strongly depends on
the differential conductivity $\sigma$ [10].

\vskip 0.5cm
\begin{center}
{\bf\S 1.4. Dynamical approximation}
\end{center}

For the small thermal $\delta T(x, t)$ and electromagnetic field
$\delta E(x, t)$
\begin{equation}
\delta T=\Theta(x)\exp[\gamma t],
\end{equation}
\begin{equation}
\delta E=\epsilon(x)\exp[\gamma t].
\end{equation}
perturbations the system of differential equations (6) and (7) can
be written in the following form

\begin{equation}
\nu\gamma\Theta=\kappa\dsf{d^2\Theta}{dx^2}+j_c\epsilon,
\end{equation}
\begin{equation}
\dsf{d^2\epsilon}{dx^2}=\dsf{4\pi}{c^2}\gamma\left[\dsf{j_c}{nE_b}\epsilon-\dsf{j_c}{T_c-T_0}\Theta\right].
\end{equation}
where $\gamma$ is the eigenvalue of the problem to be determined.
It is clear that the rate $\gamma$ characterizes the time
development of the instability. In the case, when Re$\gamma\geq
0$, small thermal and electromagnetic perturbations increase and
the stability margin corresponds to the case when $\gamma$=0. It
should be noted that the nonlinear diffusion-type equations (37)
and (38), totally determine the problem of the space-time
distribution of the temperature and electromagnetic field profiles
in the flux creep regime with a nonlinear current-voltage
characteristics in the semi-infinite sample.

As we have mentioned above, the differential conductivity
$\sigma(E)$, which determines the dynamics of the instability is
high in the flux creep regime and the parameter $\tau$ is high
enough, also [10, 11]. It is clear that this picture for the flux
jumps corresponds to the limiting case $\tau\gg 1$. Consequently,
it can be assumed that the initial rapid heating stage of a flux
jump takes place on the background of a "frozen-in" magnetic flux.
Therefore, under this dynamic approximation, we obtain from (38)
the relation between electric field $\epsilon(x, t)$ and
temperature $\Theta(x, t)$ perturbations in the following form

\begin{equation}
\dsf{j_c}{nE}\epsilon-\dsf{j_c}{T_c-T_0}\Theta=0
\end{equation}
Upon substituting the expression (39) into the equation (37) one
can easily get a differential equation for the temperature
distribution, which is conveniently presented in the following
dimensionless form

\begin{equation}
\dsf{d^2\Theta}{d\rho^2}- \rho\Theta=0.
\end{equation}
Here we introduced the following dimensionless variables
$$
\rho=\dsf{\gamma-z}{r}, \quad z=\dsf{x}{L},\quad
\dsf{1}{r}=\left[n\dsf{a L^2}{\kappa}E_{L}\right]^{1/3},\quad
E_L\sim \dot{B_e}.
$$
Thus, the condition of existence of a non-trivial solutions of
equation (40) allows to determine the spectrum of eigenvalues of
$\gamma$ and the instability threshold, accordingly. The equation
(40) has an exact solution in terms of Airy functions given as the
following form

\begin{equation}
\Theta (\rho)=c_1(s)Ai(\rho)+c_2(s)Bi(\rho).
\end{equation}
where $Ai(\rho)$ and $Bi(\rho)$ are the Airy functions. Here
constants of integration $c_1$ and $c_2$ are determined from the
thermal boundary conditions. Substituting the last solution (41)
into the thermal boundary conditions we find that $c_2=0$ and
$\Theta (\rho)=c_1(s)Ai(\rho)$. Applying the second boundary
condition we get an equation to determine the eigenvalues of the
problem

$$
\textcolor{black}{J_{2/3}(a_n)=J_{-2/3}(a_n)}.
$$
where $a_n$ are the zeros of the Bessel function, which grows with
increasing n. For example, for n=1 the stability criterion is
presented as
\begin{equation}
a_1=r^{2/3}\gamma.
\end{equation}
Using the value for the magnetic field penetration depth, we can
easily obtain from (42) an expression for the threshold magnetic
field $B_j$ at which the flux jump occurs
\begin{equation}
B_j=\dsf{4\pi j_c}{c}\sqrt{\dsf{\kappa}{anE_{L}}}.
\end{equation}
Let us now estimate the threshold field for a typical values of
parameters $j_c\simeq 10^9 A/m^2$, $T_c-T_0\simeq$ 10 K,
$\kappa\simeq 10^{-1} W/K m$, n=10. The background electric field
$E_L\simeq \dot{B_e} L$, induced by the magnetic-field variation
$\dot{B_e}\simeq 10^{-2}\div 10^{-3}$ T/s is of the order of
$E_L=10^{-4}\div 10^{-5}$ V/m for the value of $L=0.01$ m. We can
easily estimate that the threshold field has the value $B_j\asymp
1\div 3 T$.

Experimentally, the background electric field $E_L$ is created by
the sweeping rate of the applied magnetic field $\dot{B_e}$. As
can be seen from the relation (43) the threshold field $B_j$ is
decreased with the increasing of background electric field $E_L$.
It is noticeable that the dependence of the flux-jump field $B_j$
on the sweeping rate $\dot{B_e}$ of the applied magnetic field
have been verified by a numerous experiments [15-21]. An intensive
numerical analysis on the sweep rate dependence of the threshold
field has been performed recently in [22]. Recent magnetization
measurements [17] have shown that the value of the threshold field
$B_j$ decreases as the sweep rate $\dot{B_e}$ increases. A
theoretical investigations on the dependence of threshold field on
the varying external magnetic field has been performed by many
researchers (see, for example Ref. [8]). Within the framework of
the flux jump instability theory, a rapid variation of the applied
magnetic field acts as the instability-driving perturbation, and
that threshold field $B_j$ should decrease with increasing the
sweeping rate $\dot{B_e}$ [8]. The numerical studies [22] have
demonstrated that the flux jumps takes place when the sweep rate
$\dot{B_e}$ increases up to a certain value, where the number of
jumps increases with the sweep rate $\dot{B_e}$. As the sweep rate
further increases, these simulation results show that the
flux-jump field decreases and approaches a saturation value, which
is fairly close to the experimental value of about $10^2$ Oe [17].
However, as has been mentioned in [17], an experimental
investigations on the dependence of the threshold field for the
flux jump field $B_j$ on the external magnetic field sweep rate
$\dot{B_e}$ is very little. Experimentally, can be observed a
complex behavior of dependence of the threshold field $B_j$ on the
sweep rate $\dot{B_e}$. The results of experiments of Ref. [17]
demonstrated that $B_j$ is independent of the sweep rate in a
defined range of temperatures. Thus, the near independence of
$B_j$ on the sweeping rate remains to be explained. In some
conventional superconductors both the independence of $B_j$ on the
sweeping rate and its growth at a high sweeping rate [17-19] were
detected. It has been suggested [9] that a nonuniform heating may
be responsible for such an effect. However, theoretical
understanding of the thermomagnetic instabilities at such
conditions is still lacking. Note, however, that some details of
the local field behavior depend indeed on the sweeping rate, as,
for example, the number and amplitude of the jumps. In [22], it
has been demonstrated that at low values of the sweep rates
$\dot{B_e}$ the number of flux jumps decreases as sweep rate
increases. At still high sweep rates the amplitude of flux jumps
becomes independent of the sweep rate and saturates to the limit
with further increasing sweep rate.

\begin{center}
\includegraphics[width=2.5in]{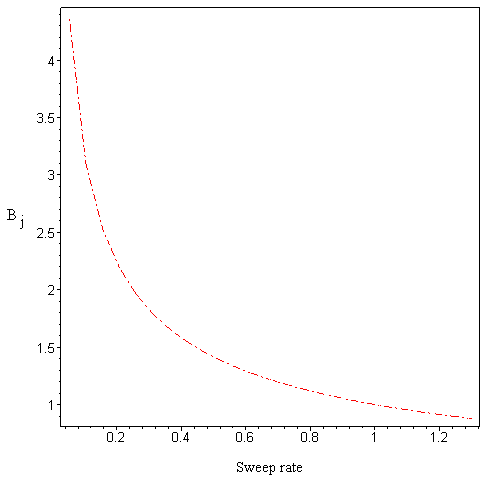}
\end{center}
\begin{center}
\end{center}
\begin{center}
\includegraphics[width=2.5in]{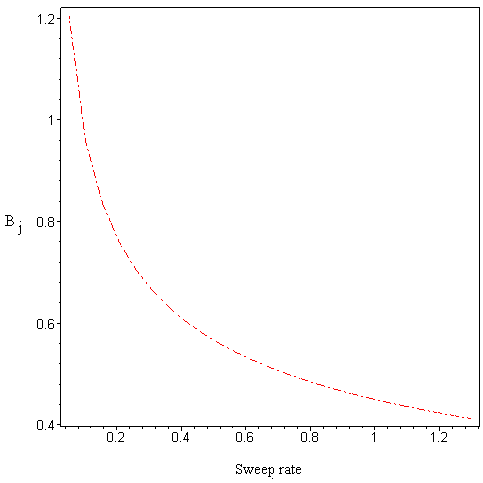}\\
\end{center}
\begin{center}
Fig. 2 and 3.  The sweep rate dependence of the threshold field
\end{center}
In Fig. 2 and 3 we have demonstrated the dependence of the
threshold field $B_j$ on the external magnetic field sweep rate at
$B_j\sim \dot{B_e}^{-1/2}$ and $\sim \dot{B_e}^{-1/3}$,
respectively. As can be seen, the value of $B_j$ decreases as the
sweep rate increases. As the sweep rate increases the value of
$B_j$ decreases and it tends to saturate at high sweep rates.
Magnetic field dependence of the critical current density only
slows down the decrease of the field $B_j$ with increasing
external magnetic field sweep rate [10, 11]. We note that for the
case Kim-Anderson model [55], the absolute value of the exponent
in the power formula decreases from 1/2 to 1/3, so $B_j\sim
\dot{B_e}^{-1/3}$ [10].

Thus, the stability criterion for the flux jumps demonstrates
extremely high sensitivity of the threshold field $B_j$ on the
values of the critical current density $j_c$, thermal conductivity
$\kappa$, and it external magnetic field sweep rate $\dot{B_e}$.
It follows from above criterion that the value of the threshold
field $B_j$ is inversely proportional to the square root of the
magnetic-field sweeping rate $\dot{B_e}$. Therefore, with the
increase of sweeping rate $\dot{B_e}$ the threshold field $B_j$
decreases, as can be seen from Figures 2 and 3.

\vskip 1cm
\begin{center}
{\bf\S 1.5.  Adiabatic approximation}
\end{center}

Let us now study the space and time evolution of small thermal and
electromagnetic perturbations within adiabatic approximation,
taking into account nonlinear flux problem. To find an analytical
solution of equations (6) and (7) we use simple adiabatic
approximation, assuming that $\tau\ll 1$. Then eliminating the
variable $\Theta(x, t)$ by using the relationship (6) and
substituting into (7), we obtain a second-order differential
equation for the distribution of small electromagnetic
perturbation $\epsilon(x, t)$ in the form

\begin{equation}
\dsf{d^2\epsilon}{dz^2}=\epsilon^{\gamma}\dsf{d\epsilon}{d\tau}-\beta
\epsilon.
\end{equation}
Here, we introduced the following dimensionless variables

$$
z=\dsf{x}{L},\quad \tau=\dsf{t}{t_0},\quad
\epsilon=\dsf{E}{E_c},\quad t_0=\dsf{4\pi}{c^2}\dsf{j_cL^2}{E_c},
\quad \gamma=\dsf{1-n}{n}.
$$
Since we have neglected the redistribution of heat in deriving
(44), only, electrodynamic boundary conditions should be imposed
on this equation

\begin{equation}
\epsilon(1, \tau)=0, \qquad \dsf{d\epsilon(0, \tau)}{dz}=0.
\end{equation}
An explicit solution of (44) together with the boundary conditions
(45) can be obtained by using the method of separation of
variables. Looking for the solution of equation (44) in the form

\begin{equation}
\epsilon(z, \tau)=\lambda(\tau)\psi(z).
\end{equation}
we get the following expressions for a new variables

\begin{equation}
\dsf{d\lambda}{d\tau}=-k\lambda^{1-\gamma},
\end{equation}
\begin{equation}
\dsf{d^2\phi}{dz^2}=k\phi^{\gamma+1}-\beta\phi.
\end{equation}
By integrating equation (47) we easily obtain

\begin{equation}
\lambda(\tau)=(\tau_p-\tau)^{-1/\gamma},
\end{equation}
where $\tau_p$ is the constant parameter, describing the
characteristic time of magnetic flux penetration profile;
$k=1/\gamma$. Now, integrating twice, the ordinary differential
equation for the function $\phi(z)$ with the boundary conditions
(45) and taking into account (49), we find the following explicit
solution for the electromagnetic field distribution

\begin{equation}
e(z,\tau)=\left[\dsf{D}{\tau_p-\tau}\cos^2\dsf{2\pi}{L^*}z\right]^{1/\gamma},
\end{equation}
$$
D=\dsf{ n^2}{1-n^2}\dsf{\beta}{2}, \quad
L^*=\dsf{1-n}{n}\dsf{\pi}{\sqrt{\beta}},
$$

The obtained solution (50) describes the distribution of the
electromagnetic field in the flux creep regime with a power-law
current-voltage characteristics. The solution describes an
blow-up-type instability in the superconductor sample. As easily
can be seen that the solution remains localized within the limited
area $x<L^*/2$ with increasing infinitively of time. In other
words, the growth of the solution, becomes infinite at a finite
time $\tau_p$. Typical distributions of the electric field
$\epsilon$(z, t) determined from analytical solution (50) is shown
in Figure 1  for the values of parameters $\tau_p$=1, n=2,
$\beta\sim$ 1 and $L^*\sim$ 0.01.

\begin{center}
\includegraphics[width=3in]{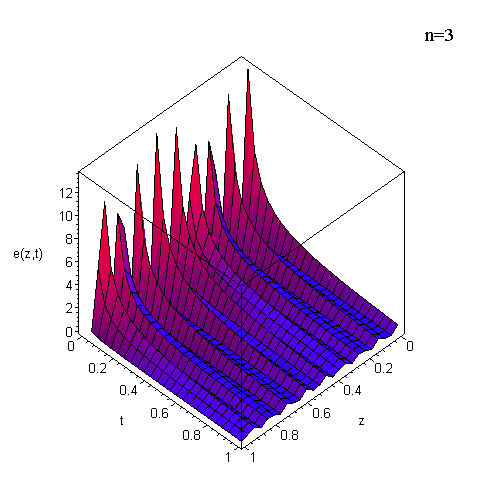}
\includegraphics[width=3in]{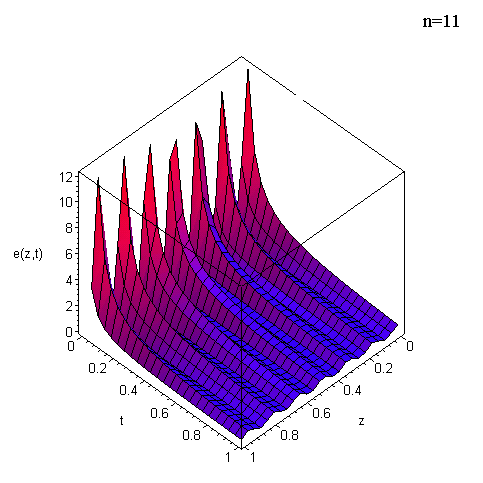}
\includegraphics[width=3in]{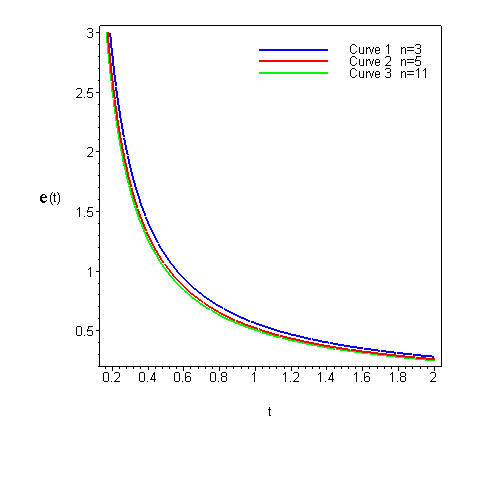}
\end{center}
\begin{center}
Fig.4a-4b. The space and time evolution of the electric field
profile at different times for  $\tau_p$=1, n=2, $\beta\sim$ 1 and
$L^*\sim$ 0.01.
\end{center}

\begin{center}
Fig.4c. The time evolution of the electric field profile at
$\tau_p$=1, n=2, $\beta\sim$ 1 and $L^*\sim$ 0.01.
\end{center}
Notice, that for the Kim-Anderson model the solution of the flux
creep problem has the form
\begin{equation}
\epsilon(z,
\tau)=\dsf{1}{\beta(\tau-\tau_0)}\left[\cos\left(\dsf{2\pi}{L}\sqrt{\beta}x\right)+1\right].
\end{equation}
Alternatively, this solution can be obtained from (50) in the
limiting case, when $n\longrightarrow \infty$.

\vskip 0.5 cm
\begin{center}
{\bf\S 1.6.  Nonuniform temperature profile}
\end{center}

It is important to notice that the profile of temperature $T(x,
t)$ may also significantly influence the condition of occurrence
of flux jumps in superconducting sample. In many cases in studying
the dynamics of thermal instabilities of superconductors were
usually assumed that the spatial distribution of temperature is
homogeneous in the cross section of the sample. However, in
reality, the temperature profile of superconductor may be
inhomogeneous along the sample as well as in its cross sectional
plane. Such inhomogeneities can appear due to different physical
reasons. First, the vortex structure can be inhomogeneous due to
existence of weak bonds in the sample. Second, the inhomogeneities
may be caused by the dependence of the critical current density on
magnetic field, the differential conductivity and the heat
conductivity. It has been recently shown [65] that under the
quasi-adiabatic conditions the temperature profile can be
essentially inhomogeneous, which strongly affects the stability
criterion of the critical state of superconductors and the first
flux jump field.

In this section we develop a phenomenological description of flux
jumps in the presence of a temperature gradient and electric field
in hard superconductors. It has been shown that thermoelectric
effects can significantly influence the stability conditions in
the mixed state of superconductors at very low temperatures.
Theoretical and experimental investigations show [66-79] that in
the mixed state during the motion of vortices inside the
superconductor sample may occur thermoelectric effects -
temperature gradients. Under certain conditions the presence of a
temperature gradient in type II superconductor sample leads to
critical state instabilities. Many investigations have been
carried out on the thermoelectric effects in type-II
superconductors, as well as in high-$T_c$ superconductors both
experimentally and theoretically [66-79]. The discovery of
high-$T_c$ superconductors stimulated intensive study of the
transport phenomena caused by the flux-flow under an electrical
current and temperature gradient. In the presence of temperature
gradient $\nabla T$ an additional thermal force appears in the
sample due to the existence of a transport entropy of vortex lines

\begin{equation}
F_t=-s\nabla T,
\end{equation}
where $s=s(T)$ is the transport entropy of a vortex line [79]

\begin{equation}
s=s_0(T/T_c)(1-T/T_c).
\end{equation}
Here $s_0$ is the entropy density neglecting the transport entropy
of the flux lines. It is notice able that the existence of the
transport entropy $s$ is associated with the presence of
low-energy electron states localized in the core of vortices and
the value of $s$ can be approximated from the local density of
states [78, 79]

\begin{equation}
s\sim \dsf{m_{e}^{3/2}\epsilon_{f}^{1/2}\xi^2}{h^3} k_{B}^{2} T.
\end{equation}
Here $\epsilon_f$ is the Fermi energy, $k_B$ is the Boltzmann
constant, $\xi$ is the coherence length.

The critical state equation with an account the transport entropy
has the form

\begin{equation}
\vec j=\vec j(T,\vec H,\vec E)+S\nabla T,
\end{equation}
where $S=\dsf{s c}{\Phi_0}$. An additional heat released due to
the transfer of transport entropy by a vortex lines, which is
proportional to $S$. Then the equation to heat flux density $\vec
q$ can be written as
\begin{equation}
\vec q=-\kappa\nabla T + \dsf{ST}{B}[\vec E,\vec B].
\end{equation}
In the quasi-stationary approximation, terms with time derivatives
can be neglected in  Eqs. (1)-(3). This means that the heat
transfer from the sample surface compensates the energy
dissipation arising in the viscous flow of magnetic flux in the
system with an effective conductivity $\sigma_f$. In this
approximation, the solution to equation (2) has the form

\begin{equation}
E=\dsf{\dot B_e}{c}(L-x).
\end{equation}
Upon substituting the expression (55) and (57) into (6), we get an
inhomogeneous equation for the temperature distribution

\begin{equation}
\dsf{d^2\Theta}{d\rho^2}-\mu\rho\dsf{d\Theta}{d\rho}-\rho\Theta=f(\rho).
\end{equation}
Here we introduced the following dimensionless variables
$$
f(\rho)=-[1+r\alpha\rho]\dsf{j_c}{aT_0},\quad
\Theta=\dsf{T-T_0}{T_c-T_0},\quad \rho=\dsf{L-x}{r}, \quad \omega
=\dsf{\sigma_f \dot B_e}{cj_c},
$$
$$
\mu =\dsf{s_0\dot B_e L^2 r^2}{c\kappa},\quad
r=\left[\dsf{c\kappa}{a\dot B_eL^2}\right]^{1/3}.
$$
Taking into account the thermal boundary conditions (8) an exact
solution to equation (58) is obtained
\begin{equation} \Theta (y) =
exp\left[\dsf{y^2}{4}-\dsf{\rho}{\mu}\right] \left[C_1D_\eta
(y)+C_2D_{-\eta-1}(y)+\Theta_0(y)\right],
\end{equation}
$$
\Theta_0(y) = -\dsf{2\pi}{\Gamma(-\eta)}D_\eta (y)\int_{0}^{y}
f(y')y'D_{-\eta-1}(y')dy'
$$
$$
+\dsf{2\pi}{\Gamma(-\eta)}D_{-\eta-1}(y)\int_{0}^{y}
f(y')y'D_{\eta}(y')dy',
$$
$$
y=\mu^{1/2}\left[\rho+\dsf{2}{\mu^2}\right], \quad \eta=\mu^{-3},
$$
where $C_1$ and $C_2$ are integration constants, which are
determined with the help of the thermal boundary conditions. In
order to estimate the maximum temperature in the sample we present
a solution to (58) in the form

$$
\Theta(x)=\Theta_m-\rho_0\dsf{(x-x_m)^2}{2}.
$$
near the point at which the temperature is a maximum, $x=x_m$.
Substituting the last relation into equation (58) we obtain an
expression for the maximum heating due to magnetic flux jumps

\begin{equation}
\Theta_m=\dsf{\left[j_c+\dsf{\sigma_f \dot B_e}{c}(L-x_m)\right]
\dsf{\dot B_e}{c \kappa T_0} (L-x_m)}{\dsf{\gamma}{L^2}-\dsf{\dot
B_e}{c\kappa}(L-x_m)
\left[\dsf{s_0\gamma}{2L^2}(2x_m-L)+a\right]},
\end{equation}
where $\gamma\sim 1$. The calculation of the maximum heating is
demonstrated that in the isothermal case, when the sample surface
is cooled intensively $w=w_0 L/\kappa\gg 1$ it is easy to verify
that the maximum heating is considerably small $\Theta_m \ll 1$
for the typical values of physical parameters. In the case of weak
sample cooling $w\ll 1$ the maximum heating is about
$\Theta_m\approx 0,5\div 2$. On the other hand, the analysis of
solution (58) show that their contribution is small in the region
of $x_m\leq x\leq L$ due to the of small amplitude of the
background electric field $E=\dot B_e(x-x_m)/c$. On the other
hand, near the point $x=x_m$ the temperature gradient $dT/dx$ is
comparatively small. Therefore, thermoelectric effects can
significantly change the temperature profile in the critical state
of the sample at the surface layer $0\ge x\ge x_m$ only, where the
background temperature gradients and the background electric field
are greater. Near the point of $x=x_m=L/2$ thermoelectric effects
are absent. The dimensionless transport entropy parameter $\mu$
can be presented in the form

\begin{equation}
\mu=\dsf{s_0}{aL}\left[\dsf{aB_{e}^{2}}{4\pi\nu j_c} \dsf{\dot
B_et_\kappa}{B_e}\right]^{1/3}.
\end{equation}
It is easily seen that $\mu\sim 1$ near the threshold for a flux
jump, when $\dsf{aB_{e}^{2}}{4\pi\nu j_c}\sim 1$ and $\dsf{\dot
B_et_\kappa}{B_e}\sim 1$. The temperature rise during the flux
jump depends mainly on the sweep rate of the external magnetic
field and heat capacity and thermal conductivity parameters in the
sample. As seen the temperature dependence $\mu(T)$ is determined
by the relationship $t_\kappa=\dsf{\nu L^2}{\kappa(T)}$. At low
temperatures the transport entropy increases with temperature as

\begin{equation}
\mu(T)\sim s_0 T^{-1/3},
\end{equation}
because $\kappa\sim T$. This means that the influence of the
thermoelectric effects on characteristics of the critical state
stability in the sample can be significant under extremely low
temperatures ($T\le 0,1 $ K). Unfortunately, quantitative data for
the transport entropy coefficient $s_0$ are available only in
limited temperature intervals thus the exact numerical evaluation
of $\mu$ proves to be difficult.

Let us investigate the stability of the critical state with
respect to small thermal $\delta T$ and electromagnetic $\delta E$
fluctuations with an account thermoelectric effects in the
quasi-stationary approximation. We present solutions to Eqs.
(1)-(4) in the form

\begin{equation}
\begin{array}{l}
T(x,t)=T(x)+\exp\left\{\ds\dsf{\lambda t}{t_{\kappa}}\right\}
\Theta\left(\dsf{x}{L}\right)\,,\\
\quad \\
E(x,t)=E(x)+\exp\left\{\ds\dsf{\lambda t}{t_{\kappa}}\right\}
\epsilon\left(\dsf{x}{l}\right)\,,
\end{array}
\end{equation}
where $T(x)$ and $E(x)$ are solutions to the unperturbed equations
obtained in the quasi-stationary approximation describing the
background distributions of temperature and electric field in the
sample and $\lambda$ is the eigenvalue of the problem to be
determined. The instability region is determined by the condition
that Re${\lambda}\ge 0$. From solution (63), one can see that the
characteristic time of thermal and electromagnetic perturbations
$t_j$ is of the order of $t_\kappa/\lambda$. Linearizing Eqs.
(6)-(7) for small $\Theta/T(x), \epsilon/E(x)\ll 1$ perturbations
we obtain the following equations in the quasi-stationary
approximation

\begin{equation}
\begin{array}{l}
\nu\lambda\Theta=\kappa\dsf{d^2\Theta}{dx^2}
+s_0E\dsf{d\Theta}{dx}
+\left[j+\sigma_{f}E\right]\epsilon-aE\Theta\,,\\
\quad \\
\dsf{d^2\epsilon}{dx^2}=\dsf{4\pi\lambda}{c^2}
\left[\sigma_f\epsilon - a\Theta+s_0\dsf{d\Theta}{dx}\right]\,.
\end{array}
\end{equation}
Eliminating the variable $\epsilon$ between two equations in (64),
we obtain a fourth-order differential equation for the temperature
distribution

$$
\dsf{d^4\Theta}{dz^4}-
\left[2\dsf{1}{f}\dsf{df}{dz}-\eta\right]\dsf{d^3\Theta}{dz^3}-
[\lambda(1+\tau)+\phi-2\dsf{d\eta}{dz}+ 2\dsf{1}{f}\dsf{df}{dz}
\left(\eta-\dsf{df}{dz}\right)+
$$
$$
\dsf{1}{f}\dsf{d^2f}{dz^2}] \dsf{d^2\Theta}{dz^2}-
[2\dsf{d\phi}{dz}+2\dsf{1}{f}\dsf{df}{dz}
\left(\lambda+\phi+\dsf{d\eta}{dz}\right)-\eta\dsf{1}{f}\dsf{d^2f}{dz^2}-
2\eta\left(\dsf{1}{f}\dsf{df}{dz}\right)^2+
$$
$$
\lambda\tau\left[\eta-\dsf{1}{\sigma_f
E}\dsf{d\eta}{dz}\dsf{df}{dz}\right]\left.\right]\dsf{d\Theta}{dz}-
[2\left(\dsf{1}{f}\dsf{df}{dz}\right)^2(\lambda+\phi)-
\dsf{1}{f}\dsf{d^2f}{dz^2}(\lambda+\phi)-
$$
$$
2\dsf{1}{f}\dsf{df}{dz}\dsf{d\phi}{dz}+ \left[\dsf{f\phi}{\sigma_f
E}-\lambda-\phi\right]]\Theta=0.
$$
Here, we introduced the following dimensionless variables and
parameters

$$
z=\dsf{x}{L}, \quad f(z)=j(z)+\sigma_f E(z), \quad
\phi(z)=\dsf{E(z)}{E_\kappa}, \quad \eta(z)=\dsf{E(z)}{E_\eta},
$$
$$.
\nu=\nu_0\left(\dsf{T}{T_0}\right)^3, \quad
\kappa=\kappa_0\left(\dsf{T}{T_0}\right), \quad
E_{\eta}=\dsf{\kappa}{s_0L^2}.
$$
Let us first consider the development of thermomagnetic
instability in the adiabatic approximation, which is valid for
hard superconductors with $\tau\ll 1$. In this limiting case, as
seen from (63), the characteristic times $t_j$ of temperature and
electromagnetic field perturbations have to satisfy the
inequalities $t_\kappa \gg t_j\gg t_m$ and $\lambda\tau\ll 1$ or
$\lambda\gg 1$. Using this approximation we reduce the last
equation to a second order differential equation

\begin{equation} \dsf{d^2\Theta}{dz^2}+
2\dsf{1}{f}\dsf{df}{dz}\dsf{d\Theta}{dz}+
\left[\left[2\left(\dsf{1}{f}\dsf{df}{dz}\right)^2-
2\dsf{1}{f}\dsf{d^2f}{dz^2}\right]
\left[1+\dsf{\phi}{\lambda}\right]-\lambda\tau\right]\Theta=0\,.
\end{equation}
Using the substitution

\begin{equation}
y=\ds\int\limits_{0}^{z}f^2(z)dz,
\end{equation}
equation (65) can be rewritten in the following form

\begin{equation}
\dsf{d^2\Theta}{dy^2}-
\left(\dsf{1}{f}\dsf{d^2f}{dy^2}\left[1+\dsf{\phi}{\lambda}\right]-
\dsf{\lambda\tau}{f^4}\right)\Theta=0.
\end{equation}
Multiplying equation (67) by $\Theta$ and integrating the result
with respect to $y$ over the interval
$$
0\le y\le y_1=\dsf{1}{L}\ds\int_{0}^{y_1} f^2(y)dy,
$$
we easily obtain

$$
\lambda\tau\ds\int_{0}^{y_1}\dsf{1}{f^4}\Theta^2dy-
\ds\int_{0}^{y_1}\dsf{1}{f}\dsf{d^2f}{dy^2}\Theta^2 dy-
\dsf{1}{\lambda}\ds\int_{0}^{y_1}\dsf{\phi}{f}\dsf{d^2f}{dy^2}\Theta^2dy=
$$
\begin{equation}
= \ds\int_{0}^{y_1}\left(\dsf{d\Theta}{dy}\right)^2dy
\end{equation}
where we have used the equality
\begin{equation}
\ds\int\limits_{0}^{y_1}\dsf{d^2\Theta}{dy^2}\Theta dy
=\Theta(y)\left(\dsf{d\Theta}{dy}\right)-
\ds\int\limits_{0}^{y_1}\left(\dsf{d\Theta}{dy}\right)^2dy
=-\ds\int\limits_{0}^{y_1}\left(\dsf{d\Theta}{dy}\right)^2dy
\end{equation}
and boundary conditions. The right-hand side of Eq. (68) has a
minimum at $\lambda=\lambda_c$:

\begin{equation}
\lambda_c=\sqrt{\tau}
\left[\dsf{\left|\dsf{1}{f}\dsf{d^2f}{dy^2}\right|_{min}}{|f^4|_{max}}
\right]^{1/2}_{min}.
\end{equation}
Now we try, to obtain an integral estimation of the increment
$\lambda$, upper and lower limits of its occurrence. The behavior
of the integrand in (70) is basically determined by the factor
$E=\dsf{\dot B_e L}{c}(1-z)$, which is equal to zero at $z=1$ (the
other factors change more smoothly). Hence, the integrand reaches
its maximum at $z=0$ and the upper estimate for $\lambda_c$ is

$$
\lambda_c\le f[j(0), T(0), E(0)]\times 1.
$$
It is evident that $\lambda_c\gg 1$ and $\lambda_c\tau\ll 1$ at
$\tau\ll 1$. Numerical evaluation gives $\lambda_c \approx 10\div
10^{2}$ at $\tau=10^{-3}$. Equations (68) and (71) enable one to
write the instability occurrence criterion in the form

\begin{equation}
\int\limits_0^{y_1} \dsf{1}{f}\dsf{d^2f}{dy^2}n_{T}^{2}dy \ge
\dsf{\pi^2}{y_{1}^{2}}+2\sqrt{\tau}
\left[\dsf{\left|\dsf{1}{f}\dsf{d^2f}{dy^2}\right|_{min}}{|f^4|_{max}}
\right]^{1/2}_{min}.
\end{equation}
Here we introduced a normalized unit vector
\begin{equation}
n_T^2=\dsf{\Theta^2}{\ds\int\limits_{0}^{y_1}\Theta^2dy}\,, \qquad
\ds\int\limits_{0}^{y_1}n_ T^2dy=1
\end{equation}
and used the inequality

\begin{equation}
\ds\int\limits_{0}^{y_1}
\ds\left(\dsf{d\Theta}{dy}\right)^2dy\geq\dsf{\pi^2}{4}\ds\int\limits_{0}^{y_1}\Theta^2dy
\end{equation}
which has been obtained with the help expansion of the function
$\Theta(y)$ in Fourier's series

\begin{equation}
\Theta(y)=A_m \cos{\dsf{y(2m+1)\pi}{2}}\,.
\end{equation}
As can bee easily seen that the shape of inequality (71)
essentially depends on the type of coordinate dependence of the
functions $j(y)$ and $f^{-1}(y)$. On the other hand, terms which
is proportional $\sqrt{\tau}$ in (70) can be as significantly
small in the limit $\tau\ll 1$. So, taking into account (74) and
above mentioned suggestions, the inequality (71) can be rewritten
in the form of

\begin{equation} \int\limits_0^{y_1}
\dsf{1}{f}\dsf{d^2f}{dy^2}n_{T}^{2}dy \ge \dsf{\pi^2}{y_{1}^{2}}.
\end{equation}
The last inequality it may be intensify to the left-hand by using
the relationship

$$
\begin{array}{l}
\left[\int\limits_0^{y_1}\dsf{1}{f}\dsf{d^2f}{dy^2}n_{T}^{2}dy\right]\le
\left[\int\limits_0^{y_1}\left|\dsf{1}{f}\dsf{d^2f}{dy^2}\right|^pdy\right]^{1/p}
\left[\int\limits_0^{y_1}\left|n_{T}^{2}\right|^{p/(p-1)}dy\right]^{(p-1)/p}=
\quad\\
\left[\int\limits_0^{y_1} \left|\dsf{1}{f}\dsf{d^2f}{dy^2}
\right|^{p}_{max}dy\right]^{1/p}\times 1= \int\limits_0^{y_1}
\dsf{1}{f}\dsf{d^2f}{dy^2}dy.
\end{array}
$$
Thus, the last stability criterion can be written in the form

\begin{equation}
\int\limits_0^{y_1} \dsf{1}{f}\dsf{d^2f}{dy^2}dy\ge
\dsf{\pi^2}{y_{1}^{2}}.
\end{equation}
The integral criterion for the critical state instability (76)
unlike the analogous criterion for a homogeneous temperature
profile, takes into account the influence of each part of the
sample and therefore temperature gradients on the threshold for
occurrence of flux jumps. Similar results can be found for the
case of composite superconductors $(\tau\gg 1)$ when instability
develops under a frozen-in conditions of magnetic diffusion.

\vskip 0.5cm
\begin{center}
{\bf\S 2. Branching flux instabilities }
\end{center}
The dynamics of vortices in type-II superconductors exhibits a
wide variety of instabilities of thermomagnetic origin. Nonuniform
magnetic flux penetration in superconductors, creating finger and
dendritic patterns, has recently attracted considerable interest.
It is generally accepted that the nonuniform penetration of the
magnetic flux is a thermomagnetic effect due to the local
overheating produced by the dissipative motion of vortices. As a
consequence of the increased local temperature, the pinning
barrier is lowered, leading to a large-scale flux invasion and to
a final nonuniform magnetic flux distribution. Such dendritic type
patterns driven by the flux jumping instability have been directly
observed under a wide variety of conditions in a large number of
superconducting samples by Duran et al. [80] and Vlasko-Vlasov et
al. [81] in Nb films, by Johansen et al. [82], Bobyl et al. [83],
Barkov et al. [84] in MgB2, Leiderer et al. [85] in other
superconducting materials by Bolz et al. [86, 87], Welling et al.
[88] and Rudnev et al. [89] by means a magneto-optical imaging
with a high spatial and temporal resolution. The existing
experimental data [80-89] and the recently developed theoretical
models [90-93], suggest that the origin of these patterns is
thermomagnetic instability of the vortex matter in the
superconducting films.

The above mentioned irregular flux avalanches and dendritic-like
patterns has a threshold applied field and temperature, when the
first avalanche occurs $B_{th}$, which strongly depends on both
temperature and the sample size, as in conventional uniform flux
jump instabilities [8-11]. The recently observed very much like
some snow-avalanches - huge compact avalanches by Welling et al.
[88] in Nb thin films by magneto-optical experiments also has a
thermomagnetic origin as proposed by Aranson et al. [91]. The
authors have determined the first flux jump field $B_j$ as a
function of temperature $T$ and found that with increasing
temperature more branching of the avalanches resulting in a more
irregular flux pattern. At higher temperatures, the number of
avalanches decreases and more flux penetration starts to dominate
the behavior. However, above certain temperature these flux
avalanches were absent. Bolz et al. [87] have studied the
dendritic flux patterns in superconducting YBCO films by means a
combination of magneto-optics and pulsed laser irradiation with
high spatial and temporal resolution in the micrometer and
nanosecond ranges. They found that dendrites only develop for
certain values of the external field and temperature. Bobyl et al.
[83] have detected in their magneto-optical experiments both
mesoscopic with the smallest size (50$\Phi_0$ at $B_e$=4 mT) and
the macroscopic of dendritic shape with the large size
($10^6\Phi_0$ at $B_e$=9 mT) flux jumps in MgB2 films. The results
of their magneto-optical experiments have shown that the jump size
increases with increasing of the applied field $B_e$. Both types
of jumps disappear above the threshold temperature (T=10 K) due to
fast increase of the specific heat and decrease of the critical
current. The authors believed that both types of jumps has a
thermomagnetic origin and its existence can be explained due to
strong demagnetization effects in thin films. Barkov et al. [84]
believed that the dentritic structures can be observed in thin
films with a sufficiently small thickness and in an adiabatic
conditions, when the heat diffusivity is much smaller than the
magnetic diffusivity. They experimentally measured a value of
threshold field $B_j$ in a superconducting thin $MgB_2$ films
using magneto-optical imaging technique which in well agreement
with the existing theoretical data [8, 9]. It is known, that
dendrite propagation in thin films shows velocities up to 160 km/s
[86], i.e., these velocities are much higher than the speed of
sound. Wertheimer and Gilchrist discovered a well defined pattern
of flux dendrites with a propagation velocity v in the interval
between 5 m/s and 100 m/s [94]. The dendrites velocity depends on
the disks thickness, for smaller d a higher velocity was found.
Bolz et al. [86]  have found that for thin superconducting films
the flux jump instability can give rise to dendritic magnetic flux
avalanches propagating with tip velocities as high as 50 km/s. The
dendritic type of flux penetration into type-II superconductor
slab was proposed by Aronson and his co-workers [91] on the basis
of numerical simulations of coupled equations for the magnetic
induction and temperature in the limit of weak Joule heating in
the sample. A formation of such dendritic flux front patterns can
be observed under condition that the magnetic flux diffusion is
much faster than the heat diffusion. They proposed that each
vortex microavalanche results in a partial flux penetration
process which is accompanied by local Joule dissipation. It has
been pointed also that the dendritic microavalanches may be also
initiated by macroscopic surface defects, which can trigger a
global flux jump instabilities in high-$T_c$ superconductors.

\vskip 0.5cm
\begin{center}
{\bf\S 2.1. Flux flow }
\end{center}
Let us consider the problem of occurrence branching instabilities
in the flux flow regime. For this purpose we formulate a basic
equation governing the dynamics of magnetic field induction and
temperature perturbation in a semi-infinitive superconducting
sample $x\geq 0$. The distribution of the magnetic flux density
$\vec B$ and transport current density $\vec j$ inside a
superconductor is given by a solution of the equation

\begin{equation}
\dsf{dB}{dx}=\mu_0j_c.
\end{equation}
The electric field $E(x, t)$ is generated inside the sample
according to Faraday's law
\begin{equation}
\dsf{dE}{dx}=\dsf{dB}{dt}.
\end{equation}
In the flux flow regime the electric field $E(x, t)$ induced by
the moving vortices is related with the local current density
$j(x, t)$ by the nonlinear Ohm's law
\begin{equation}
E=\rho (j-j_c).
\end{equation}
Let us suppose that in the flux flow regime the differential
resistivity is approximately constant and independent on magnetic
field, i.e. $\rho=\rho_f$=const [41]. The spatial and temporal
evolution of thermal T(x, t) and magnetic field B(x, t)
perturbations are described by the thermal diffusion equation
coupled to Maxwell's equations
\begin{equation}
\begin{array}{l}
\dsf{d\Theta}{dt'}=\tau\dsf{d^2\Theta}{dz^2}+ 2\dsf{db}{dz}+\Theta,\\
\quad\\
\dsf{db}{dt'}=\dsf{d^2b}{dz^2}-\dsf{d\Theta}{dz},\\
\end{array}
\end{equation}
where we have introduced the dimensionless variables and
parameters
$$
\Theta=\dsf{T}{T_c-T_0},\quad b=\dsf{B}{B_e},\quad z=\dsf{x}{l},
\quad t'=\dsf{t}{t_0},
$$
$$
l=\mu_0\rho\dsf{j_{c}B_e}{\nu},\quad
t_0=\mu_0\rho\dsf{j_{c}^{2}B_{e}^{2}}{\nu^2}.
$$
We present the small thermal and electromagnetic perturbations in
the form
\begin{equation}
\begin{array}{l}
\Theta (z, \tau)=\Theta\exp\left[\gamma t/t_0+ik\xi+iq\zeta\right],\\
\quad\\
\epsilon(z, \tau)=\epsilon\exp\left[\gamma t/t_0+ik\xi+iq\zeta\right],\\
\end{array}
\end{equation}
where $\gamma$ represents the eigenvalue of the problem to be
determined and having wave numbers $k, q=2\pi d/L$ in the
$\xi=x/L$ and $\zeta=y/L$ directions, respectively.  Solving the
system equations (80) by using the solution (81) we obtain the
following equation for the eigenvalues
$$
\gamma^2+[(1+\tau)(k^2+q^2)-1]\gamma+[\tau(k^2+q^2)-1](k^2+q^2)-2(k+q)^2=0
$$
where $\tau$ is the ratio between the characteristic time of
magnetic flux diffusion and the characteristic time of heat flux
diffusion [4]. The instability criterion for flux jumping is
determined by a positive values of the growth rate Re $\gamma$>0.
Let us first consider a simplest case, when $\tau$=0. Then, from
the last relation we obtain
\begin{equation}
\gamma^2+[(k^{2}+q^{2})-1]\gamma-(k^{2}+q^{2})-2(k+q)^2=0.
\end{equation}
First, we notice that for perturbation uniform in x-direction
$k$=0 the growth rate is positive Re$\gamma>0$ for wave number
$q<1$. In such case the small perturbations grow with the maximal
possible rate Re$\gamma=1$. For the case $q>1$, the growth rate
$\gamma$ is negative, consequently, the small perturbations always
decay. Similarly, for the perturbations uniform in y-direction
$q$=0, the growth rate is positive Re$\gamma$ only, if wave number
is satisfied $k$<1. As the wave number q approaches infinity
$q\longrightarrow \infty$, the growth rate approaches $\gamma=1$
for $k=0$ and small perturbations grow with time. It can be shown
that, for the critical values of the wave number $k=k_c$, the
growth rate is zero Re$\gamma=0$. This instability occurs at
$k<k_c$, and its criterion can be written as
$$
k<k_c=\dsf{1}{\sqrt{\tau}}.
$$
The growth rate $\gamma$ dependence on the wave number q for
different values of k are demonstrated in Figs. (5a-5d) at
$\tau=0.1$. For high enough values of wave number k the system is
stable. As the wave number $k$ decreases, the growth rate $\gamma$
increases. The branching instability will gradually appears for
relatively small values of the wave number k=0.5.

\begin{center}
\includegraphics[width=3in]{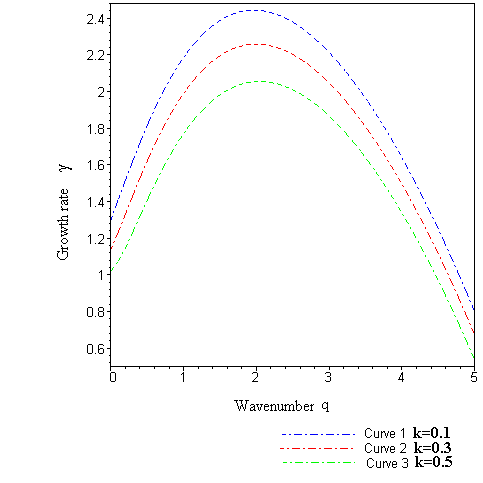}
\end{center}
\begin{center}
Fig.5a. The dependence of growth rate of the wave number k=0.5,
0.3 and 0.1 for $\tau=0.1$.
\end{center}

Thus, on the basis of a linear analysis of a set of differential
equations describing small perturbations of temperature and
electromagnetic field we found that under some conditions a
branching instability may occur in the sample.
\begin{center}
\includegraphics[width=2.5583in]{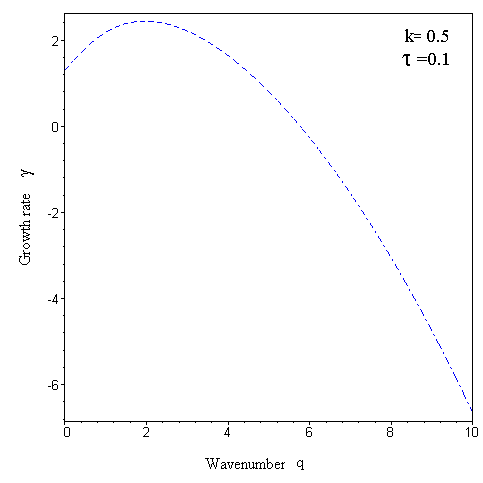}
\includegraphics[width=2.5583in]{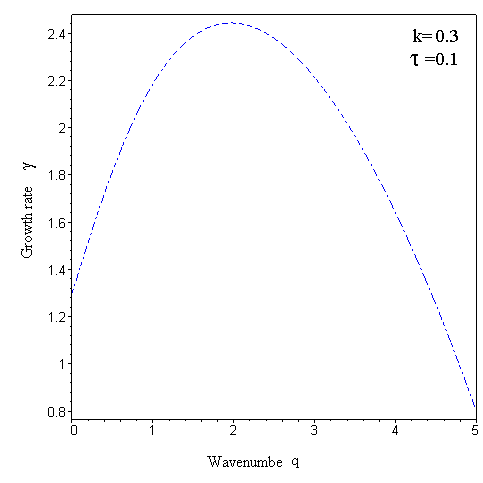}
\includegraphics[width=2.5583in]{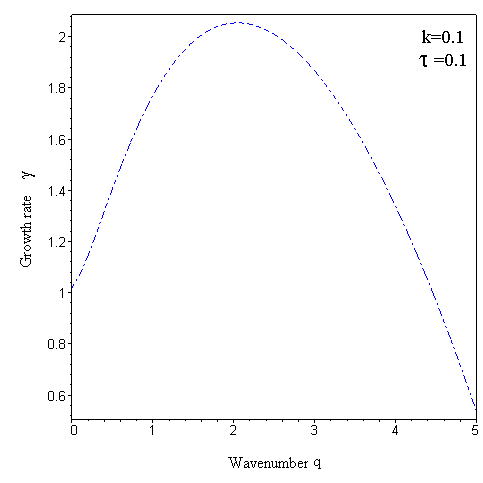}
\end{center}
\begin{center}
Fig.5b-5d.  The dependence of the growth rate on the wave number
for k=0.5, 0.3 and 0.1 for $\tau=0.1$.
\end{center}

\vskip 0.5cm
\begin{center}
{\bf\S 2.2. Flux creep}
\end{center}

Let us now study qualitatively, the problem of occurrence of the
branching instabilities within the framework of flux creep problem
with a nonlinear voltage-current characteristics. We assume that
an applied field parallel to the surface of the sample. As we have
showed above, for the flux creep problem the differential
conductivity $\sigma$ is determined by the following expression
\begin{equation}
\sigma=\dsf{dj}{dE}=\dsf{j_c}{nE_b}
\end{equation}
Taking into account this flux creep relation, we write the system
of differential equations, governing the small perturbations of
the temperature and electromagnetic field in the form
\begin{equation}
\nu\gamma\Theta=\kappa\dsf{d^2\Theta}{dx^2}+j_c\epsilon,
\end{equation}
\begin{equation}
\dsf{d^2\epsilon}{dx^2}=\mu_0\gamma\left[\dsf{j_c}{nE_b}\epsilon-\dsf{j_c}{T_c-T_0}\Theta\right].
\end{equation}
Next, we shall present a solution of these differential equations
for $\Theta(x, t)$ and electromagnetic field $\epsilon(x, t)$
perturbations in the following form
\begin{equation}
\begin{array}{l}
\Theta(x,t)=T_0(x)+(T_c-T_0)\Theta\exp\left[\gamma
t/t_0+iqz\right],\\
\quad\\
\epsilon(x,t)=E_b(x)+E_b\epsilon\exp\left[\gamma t/t_0+iqz\right].\\
\end{array}
\end{equation}
where $T_0(x)$ and $E_b(x)$ are solutions to the unperturbed
equations obtained in the quasi-stationary approximation
describing the background distributions of temperature and
electric field in the sample. From solutions (86), one can see
that the characteristic time of thermal and electromagnetic
perturbations $t$ is of the order of $t_0/\gamma$. Here, we have
introduced the following dimensionless parameters and variables
$$
t_0=\dsf{\sigma\nu a}{j_c},\quad z=\dsf{x}{l},\quad l=\dsf{\nu
a}{\mu_0 j_c},\quad q=\dsf{\pi}{2}\dsf{l}{L}.
$$
As we mentioned above, the background temperature $T_0(x)$ is
practically uniform over the cross-section of the sample and under
this approximation we ignore its coordinate dependence. It turns
out that these simplifications have no qualitative influence on
the results but make it possible to perform analytical
calculations completely.

Substituting the last expression (86) into the system equations
(84), (85) one can get the following linearized system equations
for $\Theta$ and $\epsilon$
\begin{equation}
\begin{array}{l}
\tau
q^2\Theta+\gamma\Theta+\dsf{1}{n}\Theta-2\left(\dsf{1}{n}\right)^2\epsilon=0,\\
\quad\\
q^2\epsilon+\gamma\left[\epsilon-n\Theta\right]=0,\\
\end{array}
\end{equation}
Solving the last system equations we obtain the following
dispersion relation for the eigenvalue problem
\begin{equation}
\gamma^2+\left[(1+\tau) q^2-\dsf{1}{n}\right]\gamma+\left[\tau
q^2+\dsf{1}{n}\right]q^2=0
\end{equation}
The instability of the flux front is defined by the positive value
of the rate increase Re $\gamma$>0.  It can be seen that there is
a critical cutoff wave number,
\begin{equation}
q_c=\dsf{1}{\sqrt{\tau}}.
\end{equation}
below which the system is always unstable at n=1. This instability
appears first at q=0. In this case the small perturbations grow
with the maximal possible rate $\gamma=1$. The growth rate
dependencies on the wave number for different values of $\tau$ are
demonstrated in Figs. 6a-6c at $n=1$. For high enough values of
$\tau$ the system is stable. As the  $\tau$ decreases, the growth
rate $\gamma$ increases. The branching instability will gradually
appears for relatively small values of $\tau$=0.5.

According to inequality (89), the branching instability occurs at
the threshold electric field $E=E_c$, which can be written as
$$
E_c=\dsf{\pi^2}{4}\dsf{\kappa (T_c-T_0)}{nj_cL^2}.
$$
Taking into account that, the penetration depth L, the threshold
field can be written at $B=B_{th}$  as
$$
B_{th}=\dsf{\pi}{2}\sqrt{\dsf{\kappa(T_c-T_0)j_c}{nE_b}}.
$$
The threshold field for branching instability, as can be seen from
the last expression is highly sensitive to the critical current
density and the shape of the background electric field $E_b$,
generated by the varying of magnetic field. The threshold field
$B_j$ decreases monotonously with increasing the background
electric field $E_b$. If we assume that the background electric
field $E_b$ generated by a varying of magnetic field as
$E_b\simeq\dot{B_e}$, for the considered simple geometry, then we
can easily obtain the expression for the sweep rate dependence of
the threshold field of branching instability.

Let us assume that the thermal diffusion is slower than the
magnetic diffusion $\tau\ll 1$. In this limiting case, the
instability criterion is determined as $q=q_c=1$ for n=1, so the
threshold field can be presented as
$$
B_{th}=\dsf{\pi}{2}\sqrt{\dsf{\nu}{\mu_0}(T_c-T_0)}.
$$
This is a well known adiabatic criteria [2], which assumes that
the heat transport from the sample surface to the environment can
be neglected.

\begin{center}
\includegraphics[width=2.5583in]{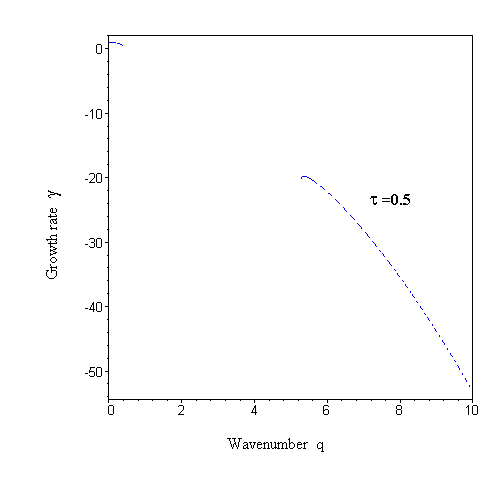}
\includegraphics[width=2.5583in]{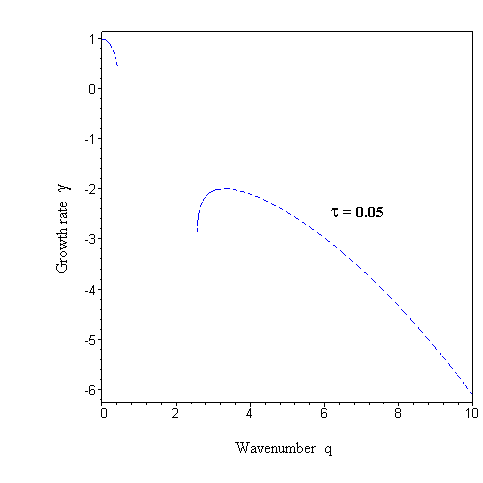}
\includegraphics[width=2.5583in]{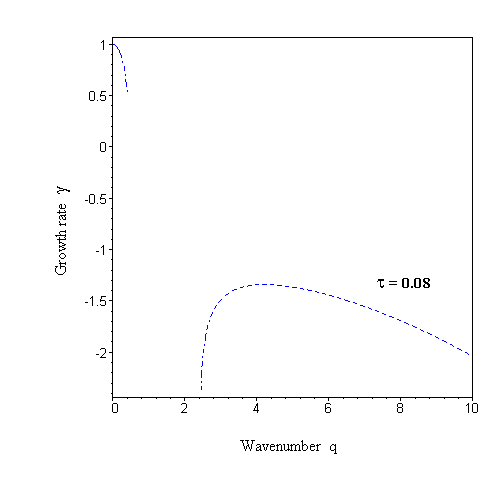}
\end{center}
\begin{center}
Fig.6a-6c.  The dependence of the growth rate on the wave number
for $\tau=0.5, 0.05, 0.08$.
\end{center}

Fingering-like patterns were found theoretically recently by
Rakhmanov et al. [90] on the linear analysis of a set of thermal
and Maxwell equations for small temperature and electric field
perturbations within the framework of adiabatic approximation.
They calculated the stability criterion and estimated the build-up
time and finger width. The fingering instability occurs at larger
electric fields $E\geq E_c$, only. The authors believed that such
fingering instability may develop to dendritic flux penetration
into the sample. By solving the Maxwell and the thermal diffusion
equations, it was shown that for small fields there are no
solutions for perturbations growing in time, implying a stable
situation. As the field increases the distribution can become
unstable, with a fastest growing perturbation having a non-zero
wave vector along the film edge. This means an instability will
develop in the form of narrow fingers perpendicular to the edge –
a scenario closely resembling the observed dendritic flux
behavior.

\vskip 0.5cm
\begin{center}
{\bf\S 3. Dynamically driven flux instabilities}
\end{center}

Recently, a many small in magnitude - mesoscopic in nature flux
jumps has been observed in magnetization measurements of
conventional and high-$T_c$ superconductors at very low
temperatures. It is interesting that the appearance of such type
flux jumps can not be described in the framework of the
traditional model of magnetothermal instabilities [1-9]. The main
features of observed flux avalanches are the following: the first
flux jump occurs at a values of the applied field much higher than
full penetration field $B_p$; the observed avalanches of almost
the same amplitude look quite regular, they are essentially
magnetic history sensitive; the avalanches exist only above
threshold field of several tesla in magnitude as opposed to the
low-field limit; its onset is independent on the sweep rate of the
external magnetic field, a distinct contrast to flux jumps
originating in a thermal instability in type-II superconductors;
the avalanche size distribution is strongly peaked with a
characteristic size, clearly distinct from the broad power-laws in
the self-organized criticality concept; the jumps occur at
significantly different fields on the increasing and decreasing
branches of the hysteresis loop. Thus, it can be concluded that
the onset of macroscopic flux jumps in these experiments may be
driven by a dynamic instability, which may be explained by means
of a concept of self-organized criticality.

For example, Seidler et al. [95] argued for a dynamical origin of
small vortex avalanches observed in their magnetization
experiments and proposed an explanation based on self-organized
criticality (SOC) [96, 97]. A new type of macroscopic flux jump
with a narrow size distribution was observed by Zieve et al. [98]
in untwined single crystals at very low temperatures (well below
$T\sim$1K), one not triggered by thermal instabilities.
Interestingly, the avalanches exist only above threshold field and
its onset is independent on the sweep rate. These avalanches occur
at significantly different fields on the increasing and decreasing
branches of the hysteresis loop. The authors concluded that the
flux jumps in their experiments may be driven by a dynamic
instability, which may be explained by means of a concept of
self-organized criticality. Based on MD simulations Olson et al.
[99] have showed that the dynamical instabilities are triggered
when the external magnetic field is increased slightly, and are
thus driven by a flux gradient rather than by thermal effects. The
existence of SOC type of behavior flux avalanches recently has
been experimentally observed and numerically adopted by many
researchers [100-109]. Altshuler et al. [109] recently have
analyzed in detail an experimental results on the flux avalanches
which may be qualitatively understood by the concept of a
self-organized criticality.

Huge "catastrophic", flux-jump-like avalanches associated with
sudden movement of many vortices were observed by Behina et al.
[110] in a superconducting Nb thick film during a slow sweep of
external magnetic field ($\dot{B_e}\sim$1.1 Oe/s) by means a Hall
probes. The obtained in a series of measurements avalanche size
statistics indicated that the size distribution of these
avalanches presents a power-law behavior only a limited range - in
the small event region. The authors believed that in contrast, at
low temperatures and fields, huge avalanches which may be related
to thermal instability of the Bean state dominate the dynamic
response of the sample.

Irregular and non-periodic flux jumps were observed by Majumdar et
al. [111] in the low field magnetization measurements at different
field sweep rates in the heavy fermion superconductor samples. The
large in magnitude flux jumps were observed at the highest sweep
rates. The magnitude of these jumps decreases slowly with
decreasing of temperature below 1.7 K and the jumps are completely
absent at 1.5 K. They argued that the observed flux jumps are due
to local flux entry through the surface or geometrical barriers.
The asymmetry of the flux jumps with respect to increasing and
decreasing part of the magnetization loop clearly indicate the
existence of barriers at surface of the sample.

Flux jumps that differ qualitatively from well-known
magnetothermal instabilities have been observed by Milner [112]
and Gerber and Milner [113] in thin single crystals of high-$T_c$
superconductors in the fields up to 17 T at the temperatures down
to 0.3 K, using magnetometric, calorimetric and induction pick-up
techniques. The flux jumps start above full penetration field
$B_p$, demonstrating clear periodicity in the slowly changing
magnetic field. The magnetization hysteresis loop is
non-homogeneous, consisting of high steps, which reflect the
global redistribution of the magnetic flux in the sample. It was
found that the frequency of flux jumps is strongly temperature
dependent. Milner proposed a few variants to explain about the
origin of the low-temperature/high-field magnetization jumps
observed in their experiments. Using a Hall probe sensor Terentiev
et al. [114] have observed a quasi-periodic flux instabilities
below a certain temperature ($T\sim$3 K) in a superconducting Nb
thin films by a square lattice of Ni dots. The magnetization
measurements indicated that the Ni dot lattice exerts a crucial
influence on the appearance and nature of the instabilities. On
the analysis of magnetization curves the authors have found that
the quasi-periodic instabilities as unexpected low-temperature
matching anomalies, most probably initiated near the film edge,
where flux density is much lower than in the film interior.

\vskip 0.5cm
\begin{center}
{\bf\S 4. Flux avalanches}
\end{center}

Recently, Chabanenko et all. [26] have reported an interesting
phenomenon in their experiments  -  convergent oscillations of the
magnetic flux arising from flux jump avalanches. The authors
argued that the observed oscillations due to flux avalanches can
be interpreted as a result of the existence of a definite value of
the effective vortex mass [27]. Thus, it is necessary to take into
account collective modes, i.e., the inertial properties of the
vortices in studying the dynamics of the flux avalanches.  Prior
to the jump, the mixed state of superconductors is characterized
by nonuniformly distributed magnetic induction localized near the
surface. As a result of the avalanche, the flux rushes from either
sides of the sample towards the center [26]. Two fronts of the
penetrating flux collide in the center of the sample and, owing to
the existing vortex mass, give rise to the local surplus density
of the magnetic flux that exceeds the value of the external
magnetic field. The repulsion force in the vortex structure at the
center of the sample that have resulted from its compaction,
initiates the wave of the vortex density of the inverse direction
of propagation. Upon reaching the surface, this wave is reflected
from it. This results in the oscillations in the vortex system
[26]. The limitation of the number of oscillations observed is
caused by the existence of damping. One succeeds in observing the
oscillation of the vortex density only owing to a strong
compression of the vortex structure as a result of the giant
avalanche-flux [17, 24-26].

In this section, we study the dynamics of the magnetic flux
avalanches, which take account inertial properties of the vortex
matter.

In the flux flow regime the electric field $\vec E(r, t)$ induced
by the moving vortices is described
\begin{equation}
\vec E=\vec v\vec B.
\end{equation}
To obtain quantitative estimates, we use a classical equation of
motion of a vortex, which it can derived by integrating over the
microscopic degrees of freedom, leaving only macroscopic forces
[32]. Thus, an equation of the vortex motion under the action of
the Lorentz, pinning, and viscosity forces can be presented as

\begin{equation}
m\dsf{dV}{dt}+\eta V+F_L+F_p=0.
\end{equation}
Here $m$ is the vortex mass per unit length, $\vec
F_L=\dsf{1}{c}\vec j\vec \Phi_0$ is the Lorentz force,  $\vec
F_p=\dsf{1}{c}\vec j_c\vec \Phi_0$, $\eta$ is the flux flow
viscosity coefficient [32]. For simplicity we have neglected the
Magnus force, assuming that it is much smaller then the viscous
force (for example, for Nb see, [26]). In the absence of external
currents and fields, the Lorentz force results from currents
associated with vortices trapped in the sample.

In combining the relation (90) with Maxwell’s equation (2), we
obtain a nonlinear diffusion equation for the magnetic flux
induction $\vec B(r, t)$ in the following form
\begin{equation}
\dsf{d\vec B}{dt}=\nabla (\vec v\vec B).
\end{equation}
\begin{equation}
m\dsf{dV}{dt}+\eta V=-\dsf{1}{c}\Phi_0(j-j_c),
\end{equation}
The temperature distribution in superconductor is governed by the
heat conduction diffusion equation
\begin{equation}
\nu (T)\dsf{dT}{dt}=\nabla[\kappa(T)\nabla T]+\vec j\vec E,
\end{equation}
Let us present a solution of equations (90-94) in the form
\begin{equation}
\begin{array}{l}
T=T_0+\Theta(x, t),\\
\quad\\
B=B_e+b(x, t),\\
\quad\\
V=V_0+v(x, t)\\
\end{array}
\end{equation}
where $T_0(x)$, $B_e(x)$ and $V_0(x)$ are solutions to the
unperturbed equations, which can be obtained within a
quasi-stationary approximation. Substituting the above solution
(95) into equations (92-94) we obtain the following system of
differential equations [26]
\begin{equation}
\dsf{d\Theta}{dt}=2v-\beta\Theta,
\end{equation}
\begin{equation}
\mu\dsf{dv}{dt}+v=-\dsf{db}{dx}+\beta\Theta,
\end{equation}
\begin{equation}
\dsf{db}{dt}=\left(\dsf{db}{dx}+b\right)+\left(\dsf{dv}{dx}+v\right),
\end{equation}
where we have introduced the dimensionless variables
$$
b=\dsf{B}{B_e}=\dsf{B}{\mu_0j_cL},\quad
\Theta=\dsf{\nu\mu_0}{B_{e}^{2}}, \quad v=V\dsf{t_0}{L}.
$$
$$
z=\dsf{x}{L},\quad
\tau=\dsf{t}{t_0}=\dsf{\Phi_0}{\eta}\dsf{B_e}{\mu_0j_cL^2}t,
$$
and parameters
$$
\mu=\dsf{\Phi_0}{\mu_0\eta^2}\dsf{B_e}{2L^2}, \quad
\beta=\dsf{\mu_0j_{c}^{2}L^2}{\nu(T_c-T_0)}.
$$
Assuming that the small thermal and magnetic perturbations has
form $\Theta(x,t), b(x,t),  v(x,t)\sim\exp[\gamma t],$ where
$\gamma$ is the eigenvalue of the problem to be determined, we
obtained from the system Eqs. (96)-(98) the following dispersion
relations to determine the eigenvalue problem
$$
(\gamma+\beta)
\dsf{d^2b}{dx^2}-[(\gamma+\beta)\mu-2\beta]\dsf{db}{dx}+[(\mu+1)\gamma^2+
$$
\begin{equation}
+[(\mu-1)\beta- \mu-1]\gamma-(\mu-1)\beta]b=0
\end{equation}
An analysis of the dispersion relation shows that, the grows rate
is positive Re $\gamma$>0, if $\mu>\mu_c=2$ and any small
perturbations will grow with time. For the case when $\mu<\mu_c$,
the growth rate is negative and the small perturbations will
decay. At the critical value of $\mu=\mu_c$, the growth rate is
zero $\gamma$=0. For the specific case, where $\mu=1$ the growth
rate is determined by a stability parameter $\beta$. Thus, the
stability criterion can be written as
\begin{equation}
\beta>1.
\end{equation}
For the case, where thermal effects is negligible ($\beta=1$) we
may obtain the following dispersion relation
\begin{equation}
\dsf{d^2b}{dx^2}-\mu\dsf{db}{dx}+(\gamma-1)(\mu+1)b=0.
\end{equation}
Seeking for $b\sim\exp(ikx)$ in dispersion relation, the growth
rate $\gamma$ dependence can be obtained as a functions of wave
number k.
\begin{center}
\includegraphics[width=3in]{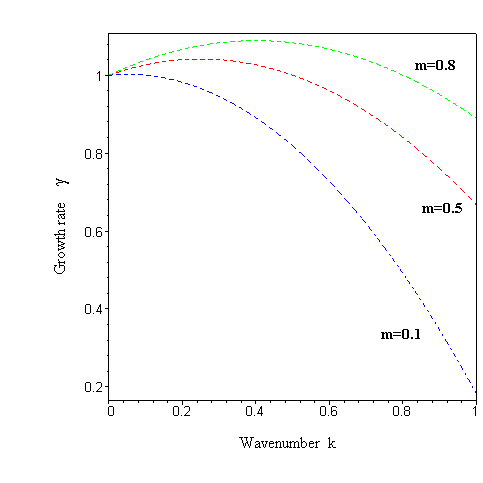}
\end{center}
\begin{center}
Fig.7a.  The dependence of the growth rate on the wave number for
$\mu=0.1, 0.5, 0.8$.
\end{center}
The stability of the system depends on the growth rate, $\gamma$,
given in (101). We analyze the growth rate of small perturbations
as a function of wave number k. When $k<k_c=\mu$ the growth rate
is positive and any small perturbations will grow with time. For
wave number $k>k_c$, the growth rate $\gamma$ is negative.
Consequently, the small perturbations always decay. It can be
shown that, for wave number $k=k_c$ the growth rate is zero
$\gamma=0$. As the wave number approaches zero $k\longrightarrow
0$ or infinity $k\longrightarrow \infty$ the growth rate
approaches $\gamma=1$ and small perturbations grow with time. As
the wave number approaches unity $k=1$ the growth rate is
determined by the value of $\mu$
$$
\gamma=\dsf{2\mu}{\mu+1}.
$$
For $\mu=0$ the growth rate is zero $\gamma=0$. For $\mu=1$ the
growth rate is unity $\gamma=1$. Since the growth rate is zero at
the critical wave number and approaches to unity in the limit of
zero wave number, there must exist a wave number in between that
maximizes the growth rate. Figs. (7a-7d) show the growing rate,
$\gamma$, as a function of the wave number k, for different values
of the parameter $\mu$. As the value of $\mu$ increases, the
corresponding growth rate increases.
\begin{center}
\includegraphics[width=2.5583in]{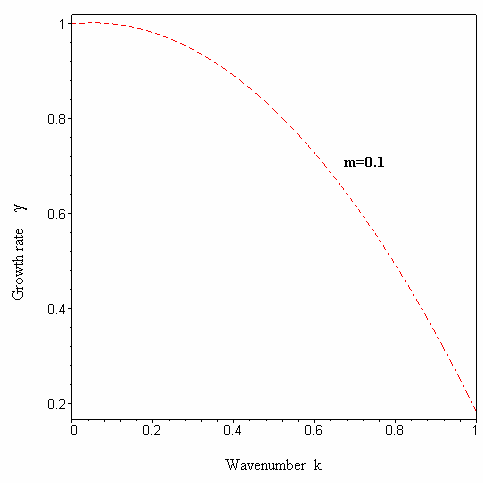}
\includegraphics[width=2.5583in]{Fig7b}
\includegraphics[width=2.5583in]{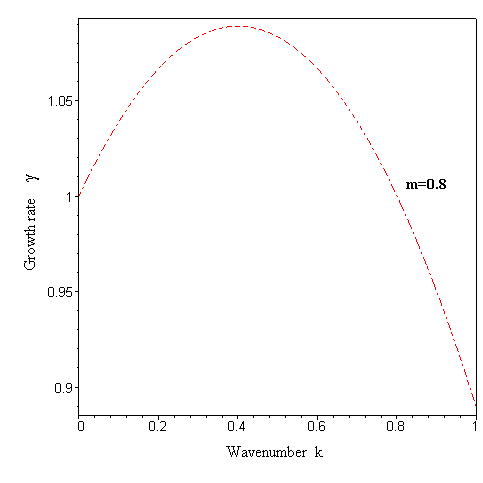}
\end{center}
\begin{center}
Fig.7b-7d.  The dependence of the growth rate on the wave number
for $\mu=0.1, 0.5, 0.8$.
\end{center}

Thus, we show that at under some conditions flux jump avalanches
may occur in superconductor sample, which take into account
inertial properties of the vortices.

\vskip 2cm
\begin{center}
{\bf\S 5. Second magnetization peak}
\end{center}

The second-peak effect in the magnetization curve is one of the
most remarkable features in the vortex state of both low and
high-$T_c$ superconductors. The second magnetization peak (or
"fishtail") as a function of applied field was observed in a
low-temperature superconductor materials [115-118] and
high-temperature superconductors [119-130]. The second peak is
expected to give an important clue in the understanding of the
complex vortex-matter phase diagram in low and high-$T_c$
superconductors. In addition, different interpretations for the
second peak have been discussed in literature, however, the
mechanism of the second peak remains unclear. These range from the
enhancement in pinning due to matching effects in oxygen deficient
structures [119], the collective creep phenomenon [120, 130],
matching effect [131, 132], the dimensional crossover [133], the
surface barrier effect [134], the thermomagnetic instability
[135-139], elastic to plastic creep [122-124], crossover in the
pinning regimes, e.g. from single-vortex pinning to a pinning of
vortex bundles [121]. Many theoretical and experimental
publications dealing with the second magnetization peak effect
have been recently published, the interpretation of this effect is
still rather controversial. No single mechanism exist until day,
which may explain the nature of these effects that were observed
in a large amount of superconducting samples.

\vskip 2cm
\begin{center}
{\bf\S 5.1. Thermomagnetic instability}
\end{center}

For low-$T_c$ superconductors (Nb), the second magnetization peak
occurs at lower temperature region during the magnetic flux
penetration, but at higher values magnetic fields close to the
upper critical field $H_{c2}$. Resent experimental investigations
indicated that second magnetization peak is probably originated by
a thermomagnetic instabilities that occur in the low-temperature
region, far below $H_{c2}$ [135-139]. In particular, such
flux-jump instabilities in Nb thin films have been studied by
Esquinazi et al. [136] on the basis of isothermal and global
magnetization measurements as a function of applied field with a
SQUID and $\mu$-Hall sensors, respectively. Large jumps as well as
pronounced irregularities of large scale gradients in the vortex
density has been observed at low temperature region ($T\sim 3$ K)
and at low applied fields ($H\sim 100$ Oe) in the magnetization
$M(H)$ curve. The existence of the second magnetization peak in a
thick Nb film, which was clearly produced by magneto-thermal
instabilities recently have been reported by Kopelevich et al.
[137, 139] using a SQUID and Hall array measurements. The
magnetization peak appears at fields higher than the first maximum
observed usually in a zero-field-cooled state, but much lower than
the upper critical magnetic field $H{c2}$. The flux jumps were
clearly observed up to the second magnetization peak at
$B_{sp}\approx 1300$ Oe. The first flux jump field $B_{fj}$ was
discussed and numerically estimated in this work. This estimation
shows that the first peak $H_{fp}\sim$ 100 Oe observed in the
magnetization curves is of the order of the first flux jump field
$B_{fj}$. Additionally, the temperature and sample size
dependencies on the magnetization jumps were investigated. It has
been shown that the maximum dissipated magnetic energy during the
flux jump process - the maximum temperature at the first flux jump
may increase substantially at $B\geq B_{fj}$ due to the decrease
of the specific heat and the increase in the critical current
density in the sample. Thus, under certain experimental
conditions, which depend on the thermal and magnetic diffusivities
and the thermal coupling of the sample with its environment, this
temperature change can produce flux jumps instabilities. It was
also shown [136] that the field of the second peak increases with
the thickness and/or width of the film as expected for
measurements in transverse geometry. The jumps become smaller in
magnitude or even vanish with a reduction of the film thickness.
The authors have pointed out that careful characterization of the
second peak in Nb films is important and may contribute to
understand the anomalous behavior of the magnetization curve
$M(H)$ measured in high-$T_c$ as well as conventional
superconductors which in some cases is interpreted in terms of a
vortex lattice phase transition [136]. Similar magnetization
measurements on the second magnetization peak were performed
recently  Stamopoulos et al. [140] at different magnetic field
sweep rates in superconductor thin films. Thermomagnetic flux jump
instability was observed at temperature T=7.2 K. The magnitude of
the first flux jump field $B_{fj}$ coincides with the first peak
$B_{pj}$ at a characteristic temperature $T=7.2$ K. The first flux
jump field remains constant $B_{fj}$ Oe at low temperature region
$T<6.4$ K. The dependence of the first flux jump field on sweep
rate of the magnetic field has been studied, also. Surprisingly
that the observed first flux jump field is not inversely
proportional to the sweep rate of the applied field, which in
contrast to theoretical concepts for the flux jumping and other
experimental dates [8, 15, 136, 17].

Let us qualitatively estimate the temperature dependence of the
first flux jump field $B_{fj}$(T), which occurs at low temperature
part of the magnetization curve. According to conclusion of Ref.
[136] the first magnetization peak $B_{fp}$(T), which is nearly
independent of the sweep rate is of the order of the field at
which the first jump $B_{fj}$(T) occurs. It is reasonable then
that the temperature dependence of the first flux jump field
$B_{fj}$(T) can be fit according to the adiabatic critical state
theory
$$
B_{fj}=\sqrt{\pi^3\dsf{\nu j_c}{-dj_c/dT}}.
$$
In order to compare the last formula with experimentally
determined first flux jump field, one must know the temperature
dependencies of the critical current density and the specific heat
of the sample. As has been mentioned in literature [136] a
quantitative estimation of $j_c$(T) from the available
experimental and theoretical models is difficult because of the
uncertainty in the values of the critical current $j_c$ at a given
field and temperature. To determine $j_c(T)$, different approaches
have been taken. Empirically, the critical current density
$j_c$(T) can be presented in the form

\begin{equation}
j_c=j_0\left[1-t^n\right]^m.
\end{equation}
where $1<m<2$. The different exponents n=1 and 2 refer to the most
common cases discussed in the literature, where the critical
current exhibits a linear and a quadratic dependence on $t=T/T_c$.

There is experimental evidence [41, 33], which indicates that the
temperature dependence of the critical current density
approximately linear at low temperatures (n=m=1). On the other
hand, at temperatures well below $T_c$ the magnitude of the
specific heat decreases rapidly with decreasing temperature, so we
can estimate the specific heat as $\nu\simeq (T/T_c)^3$. Using the
above values of temperature dependencies of the specific heat and
the critical current density we obtain the following stability
criterion [1, 2]
$$
B_{fj}=T_{c}^{2}\sqrt{\pi^3T^3(T_c-T)}.
$$
Combining all these equations, one obtains a theoretical
$B_{fj}(T)$ curve as shown in Fig. 3. Thus, it can be concluded
that the temperature dependence of the first penetration field
$B_{fp}$(T) is consistent with the temperature dependence of the
first flux jump field $B_{fp}$(T), as demonstrated in figure 8.
\begin{center}
\includegraphics[width=2.5583in]{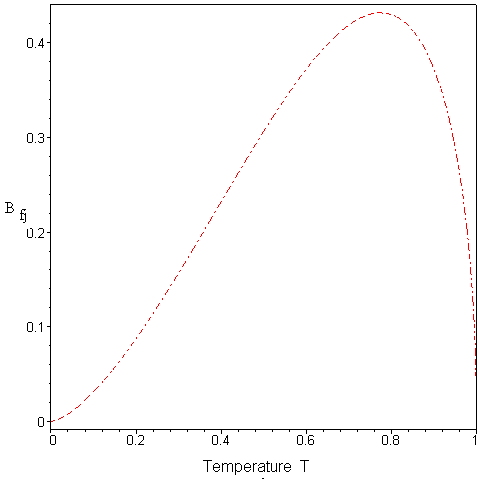}
\end{center}
\begin{center}
Fig.8. The temperature dependence of the first flux jump field
\end{center}
The dome-like profile of the temperature dependence of the first
flux jump field at low fields and temperatures have been observed
in many experiments [136, 143].  Thus, our analysis shows that the
temperature dependence of the first flux jump field $B_{fj}$(T) is
consistent with the model which ascribes the first peak (onset
field) of the second magnetization peak. This scenario is
consistent with the observation of a crossover in the field
profiles across the sample, from profiles characteristic of
geometrical barrier with weak pinning at fields below the
$B_{fp}$, to Bean-like profiles at fields above the first peak.
The results of experiments, obtained by Esquinazi [136] have shown
that near the first flux jump field the vortices penetrate the
sample, forming a droplet with a dome shape far away from the Nb
film edges, i.e., near the center or at the center of the sample,
depending on the temperature. A further increase of the external
field does not change homogeneously the flux profile inside the
sample, but part of the vortices remain pinned. The domelike flux
profiles at low enough fields have been also observed in
high-temperature platelet-shaped samples (see, for example [143]
due to the existence of a geometrical or edge barrier given by the
shape of the sample. It has been claimed [136, 138] that the
observed domelike shape of the field profile cannot be due to edge
barrier. The observed strong temperature dependence of the flux
profile at the first flux jump field $B_{fj}$(T), speaks for the
influence of a thermomagnetic instability [136]. After the first
flux jump, the increase of field moves vortices from the edges and
a Bean-like profile starts to develop. The higher the applied
field, the lower the critical current density and therefore the
larger is the penetration of the Bean-like profile into the sample
up to the second magnetization peak field $B_{sp}$.

The results of [136] indicated that the temperature dependence of
the second peak field $B_{sp}$(T) has shown a similar temperature
dependence as the upper critical field $H_{c2}$(T). So, at high
enough fields and temperatures the temperature dependence of
second magnetization peak field can be well fitted by
$$
B_{sp}(T)=B_0\left(1-t^2\right).
$$
where $B_0$ is a constant parameter. The temperature dependence of
the field $B_{sp}$(T) is shown in Fig.9.

\begin{center}
\includegraphics[width=2.5583in]{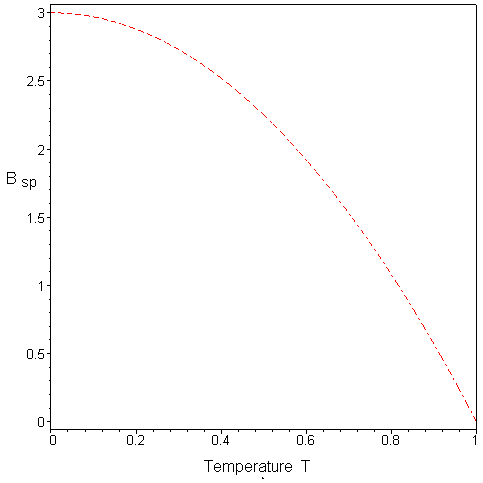}
\end{center}
\begin{center}
Fig.9. The temperature dependence of the second magnetization
peak.
\end{center}
As can be seen it reflects, roughly the temperature dependence of
the upper critical field. At higher temperatures and fields (near
the upper critical field), where the second magnetization peak
appears the temperature dependence of the critical current density
can be presented as
$$
j_c(T)=j_c(0)\left(1-t^2\right)^2.
$$
commonly accepted in literature [33] (n=m=2). Assuming that the
thermal conductivity is a linear function of temperature, we can
easily obtain an expression for the temperature dependence of the
field $B_{j}$(T)
$$
B_{j}(T)\approx B_{10}\sqrt{t\left(1-t\right)^2}.
$$
where $B_{10}$ is the free parameter.

\begin{center}
\includegraphics[width=2.5583in]{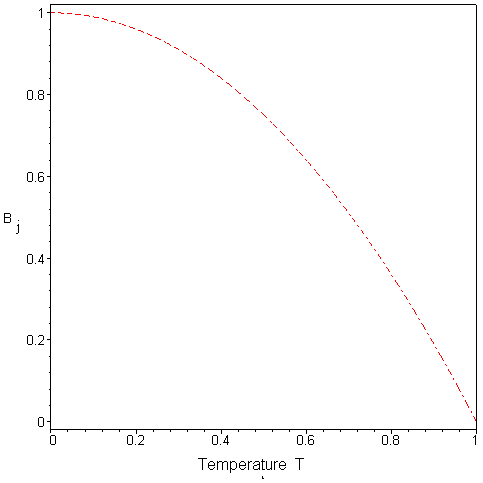}
\end{center}
\begin{center}
Fig.10. The temperature dependence of the threshold field
$B_j$(T).
\end{center}
It should be noted that the stability criterion (43) strongly
depends on the sample geometry. For the case of infinite slab
geometry or long cylindrical sample with an external magnetic
field applied parallel to its surface or axis the role of
dimension is played by the sample diameter $d=2l$. Then the
critical dimension of the sample is determined from the relation
$$
B_j=\dsf{4\pi}{c}j_c d_c,
$$
Then for the sample with a dimension d smaller than the critical
one $d<d_c$, no flux jumps occur at any temperature for any
external magnetic field. Thus by reducing the sample dimension it
is possible to avoid flux jumps. In the case of disc geometry the
stability criterion can be modified as
\begin{equation}
\beta=\dsf{\pi j_{c}^{2} r d}{\nu(T_c-T_0)},
\end{equation}
where $r$ is the radius and $d$ is the diameter of the disc. In
this case the experimental value of stability parameter for Nb
disc samples approximately is equal to $\beta$=1.2 for the typical
values of parameters $j_c$, $\nu$ and $r, d$ at helium
temperature. For thin films in an external magnetic field
perpendicular to their surface, the role of the critical dimension
from the point of view of occurrence of flux jumps is played by
their thickness. The result obtained for bulk sample may be
modified for thin samples in a perpendicular field multiplying by
a numerical factor. The dependence of the maximum in the width of
the magnetization hysteresis loop and transition field $B_{j}$ -
on the sample geometry has been reported for Nb films and single
crystals [137]. It was found that the second magnetization peak
vanishes in both superconductors for small enough dimensions of
the sample. According to flux jump theory the sample dimension is
$$
d=\dsf{c}{4\pi j_c}B_{j}.
$$
This model predicts the vanishing of the second magnetization peak
with decreasing the sample size. The estimation of the critical
thickness of the sample from the measured values of parameters and
$B_j$ show that the necessary requirement for the avoiding flux
jumping the sample thickness has to be lower than its critical
value, of the order of 100 $\mu m$ [136]. In agreement with this
model, the second peak vanishes in thin film superconductors when
the lateral sample size becomes d less than $d<d_c\simeq 100 \mu
m$. For the transverse geometry of a rectangular thin film the
effective size $d$ of the sample above which flux jump occurs is
[137]
$$
d<d_c=\left[\dsf{wl}{2}\right]^{1/2}.
$$
where $d<<l, w$; $l$ is the length, $w$ is the width of the film.
For the numerical magnitudes of a typical parameters, the critical
seise $d$ of the sample for the transpierce geometry should be
less $\leq 500 \mu m$ [137]. Hence, the authors argued that the
existence of the second magnetization peak in thick Nb film can be
should be clearly produced by thermomagnetic flux jump
instabilities [8, 9]. The authors [136] believed a such sample
size dependence of the threshold field as well as the observed
dependence of the amplitude and number of flux jumps on the field
sweeping rate $\dot{B_e}$ provide the experimental evidence for
the thermomagnetic origin of the second magnetization peak in Nb
samples. Let us estimate the temperature dependence of the
effective sample size $d_c$ for a typical experimental values of
parameters for Nb. In Fig. 11 we show the temperature dependence
of the sample size $d$. Roughly, it increases linearly with
temperature. This result was obtained from the relation for the
stability criterion (43) and it can be considered as empirical
that we assumed a linear dependence of the thermal conductivity
$\kappa=\kappa_0(T/T_c)$, where $\kappa_0$ is the constant
parameter, as suggested by low-temperature data for Nb and linear
temperature dependence of the critical current density

$$
d\approx\dsf{\pi}{2}\sqrt{t\left(1-t\right)^{-1}}.
$$
\begin{center}
\includegraphics[width=2.5583in]{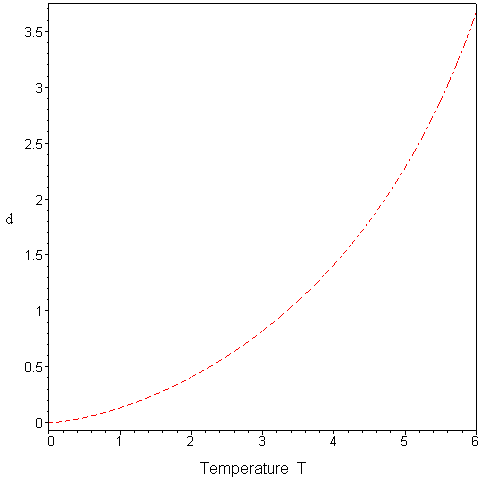}
\end{center}
\begin{center}
Fig.11. The temperature dependence of the effective sample size.
\end{center}

Nabialek et al. [17] have shown that due to the existence of a
critical dimension the flux jump instability in high-$T_c$
superconductors was observed only in relatively large single
crystals. No flux jumps has been observed in ceramic materials,
because the critical dimension in these materials is limited by
the grain size, which is typically very small. Thus, the nature of
origin and further development of the flux jump instabilities may
be different in thin films and bulk samples [93]. In bulk samples
flux jump can be triggered when the gradient of magnetic flux
profile inside the sample exceeds critical value $B_{j}$. In each
jump many vortices are involved in the avalanche process, which
normally expands to a large part of the sample volume. On the
other hand, in bulk samples, there are many large grain
boundaries, which act as strong pinning centers. The gradient of
the flux profile near these boundaries can be broken at a certain
limit. Once a blast occur, the thermal energy induced by drastic
flux motion cannot easily diffuse out and be carried by the
environment. Therefore this self-heating will lead to an increase
of temperature in the region in which the flux instability occurs,
leading to a large jump on the magnetization. One can understand
from this picture that the number of flux jumps cannot be large in
bulk samples. Such instabilities in the bulk superconductor
samples can be qualitatively understood based on the adiabatic
theory [1].  In thin films, the situation is completely different
and the observed small and irregular avalanches cannot be
explained by the adiabatic theory. There are generally more
defects or pinning centers in thin films may be observed a small
avalanches. Due to the high density of small defects formed during
the preparation process there are many places for the avalanche to
occur in thin films. Thermal diffusion is much easier in thin
films due to their very small thickness and large surface area
exposed to the environment. Therefore in thin films each avalanche
is small in magnitude but the number of avalanches can be huge.
This picture may give an explanation to many small vortex
avalanches observed in thin films. Since maintaining good thermal
contact throughout a sample is more difficult for bulk samples
than for thin films, thermally driven avalanches should occur more
easily as sample thickness increases [93].

We have studied the second magnetization peak in the framework of
a thermomagnetic model and demonstrated by a qualitative agreement
of the theory with experimental results [135-139]. In other words,
our theoretical analysis on flux jump instability qualitatively
reproduces experimentally observed in [135-139] second peak
features: the sample size dependence of the threshold field, the
temperature and sweep rate dependencies of the first flux jump
field and etc,.

Magnetic flux jump instabilities have also been observed in high
temperature superconducting materials [2, 18, 62, 19, 62, 17]. In
the following we shall discuss some recent existing theoretical
models, which may give a qualitative description on the nature of
the second magnetization peak and its onset for superconducting
materials. Both the onset field $B_{on}$ and the second peak field
$B_{sp}$ exhibit strong temperature dependence up to the close
vicinity of critical temperature $T_c$. To understand the origin
and nature of the onset of second peak it is very useful to study
the temperature dependence of the field $B_{on}$. Let us study the
temperature dependence of the onset field within the framework
above mentioned theoretical models. In the last decade, many
different by each other hypotheses have been proposed to explain
the nature of the second peak effect. For example, Krusin-Elbaum
et al. [121] ascribe the occurrence of the peak effect to the
crossover between two collective vortex creep regimes; from single
to collective vortex creep regime. The independence of the
magnetic moment from the applied field for low temperature
isothermal loops is interpreted by them as the single vortex creep
regime [121]. The authors believed that the peak effect is not
likely to be observed on very short time scales due to the fact
that the relaxation is much faster in the single vortex creep
regime than in the collective vortex creep regime. Yeshurun et al.
[130] also assigned dynamical characteristics to the appearance of
the peak effect. In other studies, the influence of oxygen content
on the peak behavior is discussed [142]. Zhukov et al. [142] found
that the peak effect was present at high temperatures for all
oxygen contents, however the peak’s height takes the smallest
value for the state closest to stoichiometric status, which means
that oxygen deficiency is necessary for achieving greater pinning
and therefore higher critical current density. In some cases a
correlation between the peak effect and the oxygen concentration
is found to be the source of peak effect. Daumeling et al. [119]
argued that a field-induced pinning enhancement in
oxygen-deficient regions may be the source of the second
magnetization peak in superconductor sample. The authors proposed
that oxygen-deficient areas had become normal as the field was
increased, as a result of their lower T and H. These areas thus
become new pinning centers. Therefore any change in oxygen content
or distribution should lead to suppression or appearance of the
peak effect. In a quite different model, the peak effect is
considered as a result of the matching effect [132], where
matching of the vortex lattice with the defect structure leads to
pinning enhancement. However, the temperature-dependent nature of
the peak strongly rules out the matching effect as the origin of
the peak effect.

The experiments of magnetic relaxation are the most extensively
used tools to study the vortex dynamics in a variety of
superconducting materials [121-124].  The relaxation rate S as a
function of H can provide relevant information about the pinning
mechanism [121]. It has been demonstrated [121] that the
relaxation rate drops as field increases from below $B_{sv}$ up to
a field close to second magnetization peak field $B_{sp}$,
increasing again as H become larger than $B_{sp}$. Results of this
magnetic relaxation experiments have shown the existence of a
crossover in the pinning mechanism at $B_{sv}$, where apparently,
this crossover occurs without a change in the behavior of the
relaxation rate with field. The peak in S(H) has been recently
interpreted [121] as a consequence of the crossover from
individual to collective regime. At the high-temperature side of
the peak in S(H), relaxation is anomalous in the sense that the
relaxation rate varies non-monotonically with  time. It has been
shown [121] that the anomalous relaxation that occurs on the
high-temperature side of the peak in S(H) indicates that the
vortex structure undergoes a change in the relaxation regime. The
second magnetization peak occurring at $B_{sp}$ is formed by a
crossover in the pinning mechanism, from collective to plastic
pinning as the field increases [122]. Results of this work show
that the temperature dependence of the second magnetization peak
position, $B_{sp}$ is well explained in terms of a plastic motion
of the vortex lattice. An analysis of the relaxation data shows
that the peak field is moving with time to lower values. The
dynamic picture is also consistent with results by Yeshurun et al.
[130], showing a clear time dependence of the second peak. Both
studies show that the peak is virtually absent when the loops are
measured on a short time scale while the peak gradually appears as
the time scale is expanded. As pointed out in the previous
section, the activation energy grows with B below the second peak
and decreases after the peak. This crossover in the field
dependence indicates an elastic-to-plastic creep crossover around
the peak-field similar to that observed in other materials
[122-130].

\vskip 2cm
\begin{center}
{\bf\S 5.2. Collective creep}
\end{center}

One of simple situation to introduce the concept of weak
collective pinning is an isolated vortex in an isotropic
superconductor subject to pinning by randomly distributed point
defects [141]. We shall study the problem within a continuum
elastic description of the flux-line lattice, with a potential
energy given by the combination of the elastic energy, the pinning
energy and the action of the Lorentz force

\begin{equation}
E_e=c_{66}u^2L+e_1\dsf{u^2}{L}-(\gamma\xi^2 L)^{1/2}.
\end{equation}
where u and L are the transverse and longitudinal sizes of vortex
distortion respectively, $\gamma=f_{p}^{2}n_i\xi$ is the disorder
parameter; $f_p$ is the elementary pinning force, $n_i$ is the
density of pinning centers, $c_{66}=e_0/4a_{0}^{2}$ is the
vortex-lattice shear modulus; $e_1=e^2e_0$ is the vortex line
tension, e is the anisotropy parameter, $e_0$ is related to the
energy of a vortex line per unit length
$e_0=(\Phi_0/4\pi\lambda)^2$, $a_0=\sqrt{\Phi_0/B}$ is the vortex
line spacing,  $\Phi_0$ is the flux quantum, $\lambda$ is the
penetration depth, $\xi$ is the coherence length. The
characteristic longitudinal length $L_0$, which determines the
size of elastic screening of local distortions of the vortex line
can be evaluated by minimizing of the elastic energy with respect
to L

$$
L_0=\sqrt{\dsf{e_1}{c_{66}}}.
$$
The characteristic elastic energy of the vortex thus becomes

\begin{equation}
E_e=\dsf{ee_0}{a_0}u^2.
\end{equation}
The collective pinning length can be evaluated by balancing the
elastic energy cost and the pinning energy gain associated with a
small displacement of a region of linear size L

\begin{equation}
L_c=\left(\dsf{e_{0}^{2}\xi^2}{\gamma}\right)^{1/3}.
\end{equation}
Equating the resulting pinning force and the Lorentz force on
segments of the length $L_c$ allows one to obtain the critical
current density for the case of weakly pinning point defects and
small applied fields

\begin{equation}
j_{sv}=\dsf{c}{\Phi_0}\left(\dsf{\gamma}{L_c}\right)^{2}.
\end{equation}
Let us estimate for the single-vortex—small-bundle crossover field
$B_{sv}$, where the system changes from single-vortex creep to
collective creep of small bundles. This crossover field $B_{sv}$
is determined by the condition [49]

$$
L_c=ea_0.
$$
This condition is equivalent to

\begin{equation}
B_{sb}=\beta\dsf{j_{sv}}{j_0}B_{c2},
\end{equation}
where $\beta\approx 5$ is the constant parameter,
$j_0=4B_c/3\sqrt{6}\mu_0\lambda$ the depairing current density,
$B_c=\Phi_0/2\sqrt{2}\pi\lambda\xi$ the thermodynamic critical
field, $B_{c2}=\mu_0\Phi_0/2\pi\xi^2$ the upper critical field.

As can be seen from the last expression, the temperature
dependence of the crossover magnetic field $B_{sv}$ is determined
by the temperature dependence of the intrinsic superconducting
parameters, such as the coherence length $\xi$(T) and the magnetic
penetration depth $\lambda$(T), and the disorder parameter
$\gamma$ of the vortex system. According to G. Blatter et al.
[49], defects can interact with the vortices in two different
ways. They may cause a spatial variation of the transition
temperature ($\delta T_c$-pinning) and a spatial modulation of the
mean free path ($\delta l$-pinning). In both cases, the influence
of disorder is described by a disorder parameter $\gamma$,
proportional to the defect density and the temperature dependence
of $\gamma$ is different for the two cases. For $\delta
T_c$-pinning, $\gamma=\lambda^{-4}$, while for $\delta l$-pinning,
$\gamma=\lambda^{-4}$. Thus, depending on the type of pinning the
crossover field can be either an ascending or descending function
of temperature. Considering the functional form of the disorder
parameter for $\delta T_c$-pinning, $\gamma=\lambda^{-4}$. Within
the frame of the Ginzburg-Landau theory
$\lambda(T)=(1-t^4)^{-1/2}$ and $\xi(T)=[(1-t^2)/(1-t^2)]^{-1/2}$,
where $\lambda(0)$ and $\xi(0)$ is the penetration depth and the
coherence length at T=0 K. Inserting the values of the coherence
length $\xi(T)$ and the penetration depth $\lambda(T)$ into the
last equation for $B_{sp}$(T) we obtain the following expression
for the crossover field $B_{sv}$ in the single creep regime for
the case $\delta T_c$ pinning

\begin{equation}
B_{sv}=B_{sv}(0)\left(\dsf{1-t^2}{1+t^2}\right)^{2/3},
\end{equation}
and  $\delta l$-pinning, respectively

\begin{equation}
B_{sv}=B_{sv}(0)\left(\dsf{1-t^2}{1+t^2}\right)^{2}.
\end{equation}
\begin{center}
\includegraphics[width=2.5583in]{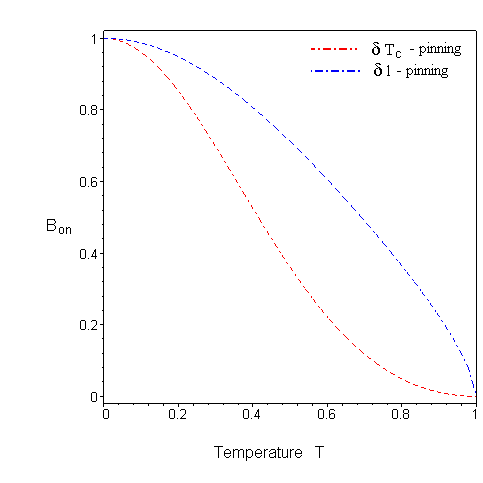}
\end{center}
\begin{center}

Fig.12. Crossover field as a function of temperature corresponding
to $\delta l$ and $\delta T_c$ pinning.
\end{center}

The data in Figure 12 clearly show that only, the $\delta
T_c$-pinning formula can explain the temperature dependence onset
field $B_{on}$, accurately. Thus, a crossover from individually
pinned regime to a collectively pinned regime is to be anticipated
at a threshold field $B_{sv}$, above which vortex-vortex
interactions become dominant. At low temperatures with the
numerical values $\lambda$ and $\xi$ we find that $B_{sv}$ is of
the order of 1.7 T. This is justified in the limiting case of
small magnetic fields. Consequently, the transverse correlation
length of the flux lattice $L_c$=1 is approximately equal to $0.1
a_0$, i.e. it is much lower than the intervortex spacing $a_0$.
Thus, the vortex system is in the single-vortex pinning regime at
low fields.

As we have already mentioned above, the analysis of collective
creep in terms of individually moving vortices applies only for a
limited regime at $B=B_{sv}$. However, with increasing applied
field the relative importance of the interaction between the
vortices grows and for larger field values, $B_{sv}\leq B$ we
enter to the collective or elastic creep, where thermal
fluctuations become large due to large flux creep [122-130]. It
was now strongly experimentally established that, the line
$B_{sp}$ can be well explained by a plastic creep model based on
dislocation mediated motion of vortices similar to diffusion of
dislocations in atomic solids [122-130]. The activation energy
$U_{pl}$ for the motion of a dislocation in the vortex lattice can
be estimated as

\begin{equation}
U_{pl}=\dsf{1}{4\pi\gamma}e_0a_0\propto B^{-1/2}.
\end{equation}
One notices that $U_{pl}$ decreases with field in contrast to the
collective creep activation energies $U_{el}$, which increases
with field. The fishtail peak location $B_{sp}$ can be determined
from the condition $U_{pl}=U_{el}$. Using the logarithmic solution
of the flux diffusion equation we find

\begin{equation}
U_{el}=kT\ln\dsf{t}{t_0}.
\end{equation}
Balance between the $U_{pl}$ and $U_{el}$ energies gives an
expression for the temperature dependence of $B_{sp}$. To roughly
estimate the crossover field and equate the values of the elastic
and plastic pinning barriers [122], the following relation can be
used

\begin{equation}
\dsf{e_0}{\gamma\sqrt{B_{sp}}}=kT\ln t.
\end{equation}
From the last relation we find
\begin{equation}
B_{sp}=\Phi_0\dsf{e^2e_{0}^{2}}{(kT)^2}\propto e_{0}^{2}.
\end{equation}
Inserting the values of the penetration depth $\lambda$ to the
equation for the $B_{sp}$ we obtain the temperature dependence of
the transition field, as it is demonstrated in Fig. 13. Note, that
a similar temperature dependence $B_{sp}$(T) was found in some
experiments [116] for the melting line. This interesting
similarity may support previous claims [117] that the plastic
motion of defects in the vortex lattice is a precursor to the
melting transition.

\begin{center}
\includegraphics[width=2.5583in]{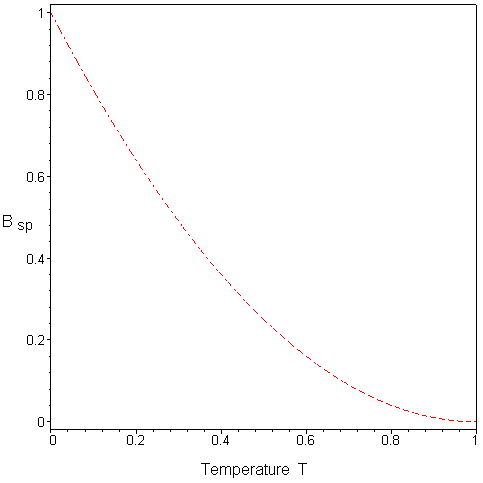}
\end{center}
\begin{center}
Fig.13. Second magnetization peak field as a function of
temperature.
\end{center}

In Fig. 13 we show the temperature dependence of the peak field.
Thus, our analysis of the vortex-dynamics shows that the onset of
the second magnetization peak occurring at $B_{on}$ is formed by a
crossover in the pinning mechanism, from single to collective
pinning as the field increases. On the other hand, the temperature
dependence of the second magnetization peak $B_{sp}$ can be well
explained by the last expression for the plastic vortex creep
model.

\vskip 0.5cm
\begin{center}
{\bf\S 5. 3. Order disorder transition}
\end{center}

The nature of the different vortex phases in the mixed state of
high temperature superconductors is a topic of extensive
experimental and theoretical research [49, 143-146]. Much of the
experimental work has been focused on the highly anisotropic
(BSCCO) crystals, revealing a rich phase diagram [143, 145, 146].
In particular, recent local magnetization measurements [143] in
BSCCO crystals revealed a sharp onset of an anomalous second
magnetization peak at a field $B_{on}$, which was interpreted as
indicating a transition between two solid phases of the vortex
structure. An evidence for two distinct solid vortex phases in
BSCCO was previously obtained in neutron diffraction [145] and mSR
[146] experiments. Following these observations, a theoretical
model was developed [147–149], describing a mechanism for a
disorder-induced transition, from a relatively ordered vortex
lattice, to a highly disordered vortex state.

An experimental studies have shown [153, 154] that the sample
exhibits a distinct second peak, i.e., a strong increase in the
magnitude of the magnetization in an intermediate field range. The
lowest of these is $B_{on}$ denoting the onset field of the second
peak while the peak maximum occurs at $B_{sp}$. At low fields the
elastic interactions govern the structure of the vortex solid,
forming a quasiordered lattice [150]. Above the onset field
$B_{on}$, disorder dominates and vortex interactions with pinning
centers result in an entangled solid where cells of the vortex
lattice are twisted and dislocations proliferate [147-149]. To
understand the origin of the onset peak it is very useful to study
the temperature dependence of the field $B_{on}$. By considering
the competition between the elastic energy of the vortex system
and the pinning energy [147–149] it has been suggested that the
second peak can result from a transition of a low-field
quasiordered vortex phase to a disordered vortex solid at higher
fields, induced by the quenched disorder. According to recent
theoretical model [147–149], the vortex phase diagram is
determined by the interplay between three energy scales: the
vortex elastic energy $E_{el}$, the energy of thermal fluctuations
$E_{th}$ and the pinning energy $E_p$. The competition between the
first two determines the melting line while the competition
between the last two determines the irreversibility line. The
competition between the elastic energy and the pinning energy
determines the order-disorder transition field $B_{on}$. We now
study the temperature dependence of the onset field, within the
framework of a theoretical model [147–149], based on the
phenomenological Lindemann criterion for the regime of single
vortex pinning [52]. The order-disorder occurs when the
disorder-induced wandering of the flux lines is comparable to the
lattice constant $a_0$ and the vortex system loses its
translational order and transforms into an disordered phase, in
which vortices better adapt to the local pinning potential. To
describe the order-disorder phase transition line, we use the
following Lindemann criterion

\begin{equation}
\langle u^2(L_0, 0)\rangle=c_{L}^{2}a_{0}^{2},
\end{equation}
where $\langle u\rangle$ describes the mean relative displacement
of neighboring flux lines caused by the disorder. This Lindemann
criterion leads to the usual condition for the order–disorder
transition at low temperatures, when the thermal fluctuation is
negligible. The Lindemann number $c_L$ is introduced here as a
phenomenological parameter that is supposed to depend only weakly
on the specific lattice parameters of the solid phase, in
particular it is assumed to be independent of the magnetic field.
Strictly speaking, the value of the constant $c_L$ may depend on
whether the order–disorder transition occurs in the single vortex
pinning region or in the region of bundle pinning. However, to
understand the essence of the matter, we shall use the simplest
approximation: $c_L$ will be considered as the same constant for
the various regimes of pinning. We, therefore, use this value in
our calculations.

The physics of a single vortex line in point disorder exhibits two
different scaling regimes depending on the typical size of the
disorder-induced mean-square displacement

\begin{equation}
\langle[u(L)-u(0)]^2\rangle=u_{c}^{2}\left[\dsf{L}{L_c}\right]^{2\zeta},
\end{equation}
of a vortex segment of length L, where <.........> denotes the
full statistical average to be taken over dynamical variables
first and then over the quenched variables describing the
disorder, $\zeta$ is so-called static critical exponent. We have
to carefully distinguish several pinning regimes depending on the
size of the pinning length $L_c$ in comparison to the
single-vortex length $L_0$. For short vortex lengths $L_0<L_c$
displacements are small
$$
\langle u^2(L_c)\rangle\simeq\xi^2,
$$
perturbation theory is valid, and we can work with Larkin’s random
forces [141]. Fluctuations around the ground state of the line are
Gaussian in this random force regime and we find a roughness
exponent $z=3/2$, i.e.,

$$
\langle[u(L)-u(0)]^2\rangle=\xi^2\left[\dsf{L}{L_c}\right]^3.
$$
This regime is valid up to the collective pinning or Larkin length
$L_c$ which is defined by the condition

$$
\langle u^2(L_c)\rangle=\xi^2.
$$
Longer segments explore many almost degenerate minima of the
pinning energy landscape such that fluctuations are non-Gaussian.
In this so-called random manifold regime the roughness exponent is
known exactly $z=5/2$ [155]. The currently most reliable estimate
for general n is $z=5/8$ for the vortex line in the bulk
superconductor. Therefore we find

$$
\langle[u(L)-u(0)]^2\rangle=u_{c}^{2}\left[\dsf{L}{L_c}\right]^{5/4}.
$$
On scales exceeding the Larkin length $L_c<L_0$, the vortices
start to explore many minima of the disorder potential [155]. This
regime is referred to as the random manifold regime [155, 151]. In
the case $L_c<L_0$ the pinning energy becomes

\begin{equation}
E_p=U_{dp}\left[\dsf{L_0}{L_c}\right]^{1/5},
\end{equation}
where
$$
U_{dp}=\left(\dsf{\delta e_0\xi^4}{\gamma^4}\right)^{1 /3}.
$$
By comparing the pinning $E_p$ and elastic energies $E_{el}$ we
now derive the transition field

\begin{equation}
B_{on}=B_0\left[\dsf{T_0}{U_{dp}}\right]^3,
\end{equation}
$$
B_0=\dsf{c\Phi_0}{\xi^2}, \quad T_0=\dsf{cee_0\xi}{2}.
$$
Each vortex remains individually pinned in the presence of thermal
fluctuations until the thermal energy T is greater than the
typical depinning energy $U_{dp}$ of each single vortex. After
further analytical calculation we obtain the following formula to
describe the temperature dependence of the transition field

\begin{equation}
B_0=\dsf {c^5ee_{0}^{2}\Phi_0}{\sqrt{2}\gamma\xi^{3}}.
\end{equation}
As we have mentioned above, depending on the type of pinning the
crossover field can be either an ascending or descending function
of temperature. Substituting the temperature dependencies of the
quantities   $\xi(T)$, $\lambda(T)$ into the relation (119)  and
considering the functional form of the disorder parameter, we
arrive at the following expressions for $\delta T_c$-pinning

\begin{equation}
B_{on}\propto (1-t^4)^{3/2}.
\end{equation}
and for $\delta l$-pinning

\begin{equation}
B_{on}\propto (1-t^4)^{-1/2}.
\end{equation}
respectively.
\begin{center}
\includegraphics[width=2.5583in]{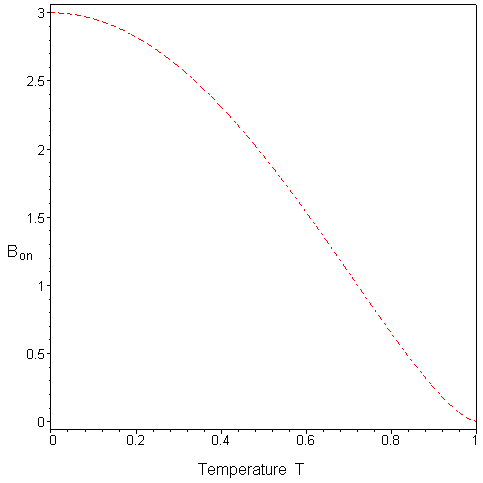}
\end{center}
\begin{center}
Fig.14. Onset field as a function of temperature.
\end{center}

In Fig. 14 we show the temperature dependence of the onset field.
The data in Figure 14 clearly shows that $B_{on}(T)$ for $\delta
T_c$ pinning is a decreasing functions of temperature and
therefore this formula can explain the onset of the second
magnetization peak, properly.

It is notice able, that onset field $B_{on}$ is inversely
proportional to both the anisotropy $\gamma$ and the disorder
parameter $\gamma$. Thus, both the anisotropy and disorder may
sufficiently affect to the shape of temperature dependencies of
both the onset and second magnetization peak field. As has been
pointed out Yeshurun et al. [130] that the point disorder induced
by electron irradiation can modify the penetration depth
$\lambda$, anisotropy $\gamma$ and critical temperature $T_c$, and
thus cause a significant shift of the transition line. Thus, the
line $B_{on}$(T) can be expected to shift lower fields with the
introduction of additional disorder in the crystal structure.
Calculations [147-149] of $B_{on}$, based on a Lindemann
criterion, confirm this expectation. Other experimental studies
Khaikovich et al. [143] of the effect of electron irradiation on
the order-disorder transition line have found a systematic
decrease of $B_{on}$(T) with increasing irradiation dose. The
decrease in $B_{on}$ is consistent with the enhanced vortex
pinning after electron irradiation. Therefore, the quasiordered
phase is stabilized by introduction of the point disorder and the
strong pinning region is expanded to the lower fields. These
results of electron irradiation effect provide further evidence of
the field-driven disordering transition scenario [147-149] as a
possible origin of $B_{on}$(T). Specifically, the decrease in the
superfluid density $n_s$ with increasing doping at high
temperatures increases the magnetic penetration length $\lambda$,
hence decreases $B_{on}\propto \lambda^{-4}$. It can be concluded
that introduction of even a small amount of disorder into very
clean systems changes the phase diagram drastically by lowering
the order-disorder transition line $B_{on}$(T) and also changing
it’s temperature dependence. When the disorder reaches a certain
threshold, however, introducing additional disorder does not
continue this tendency. With other words, in highly disordered
systems, the vortex matter phase diagram is relatively robust with
respect to variations in the exact degree of disorder [156-158].

\vskip 0.5cm
\begin{center}
{\bf   Conclusion}
\end{center}
In the present paper we studied the flux jump instabilities of the
critical state in conventional and high-$T_c$ superconductors. We
have determined the flux jump field and critical state stability
criterion within framework of Bean's model. To determine flux
jumps threshold an analytical simulations of coupled equations for
the magnetic, electric field inductions and temperature has been
performed. The field of the first flux jumps is calculated
analytically using the dynamic and the adiabatic approximations.
The qualitative analysis of the magnetic flux jumps instabilities
for superconductors is provided.

\vskip 0.5cm
\begin{center}
{\bf   Acknowledgements}
\end{center}
This study was supported by the Volkswagen Foundation Grant.

\newpage \vskip 0.5cm
\begin{center}
{\bf   References}
\end{center}

\begin{enumerate}
{\item  P. S. Swartz and S. P. Bean, J. Appl. Phys., 39, 4991,
1968.

\item  C. P. Bean, Rev. Mod. Phys., 36, 31, 1964.

\item  H. London, Phys. Lett., 6, 162, 1963.

\item  S. L. Wipf, Phys. Rev., 161, 404, 1967; L. S. Wipf,
Cryogenics 31, 936 1992.

\item H. R. Hart, J. Appl. Phys., 40, 2085, 1969.

\item  R. Hansox, Phys. Lett., 16, 208, 1965.

\item  M. G. Kremlev, Cryogenics, 14, 132, 1974.

\item R. G. Mints and A. L. Rakhmanov, Instabilities in
superconductors, Moscow, Nauka, 362, 1984.

\item  R. G. Mints and A. L. Rakhmanov, Rev. Mod. Phys., 53, 551,
1981.

\item R. G. Mints, Phys. Rev., B 53, 12311, 1996.

\item R. G. Mints and E.H. Brandt, Phys. Rev., B 54, 12421, 1996.

\item  M. M. Mola, S. Hill, J. S. Qualls and J. S. Brooks, Int. J.
Mod. Phys., B 15, 3353, 2001.

\item  H. A. Radovan and R. J. Zieve, Phys. Rev. B 68, 224509,
2003.

\item  E. R. Novak, O. W. Taylor, Li Liu, H. M. Jaeger, and T. I.
Selinder, Phys. Rev., B 55, 11702, 1997.

\item  S. Khene and B. Barbara, Solid State Comm., 109, 727, 1999.

\item M. N. Wilson, Superconducting Magnet, Oxford University
Press, London, 131, 1986.

\item   A. Nabialek, M. Niewdzas, H. Dabkowski, J. P. Castellan,
and B. D. Gaulin, Phys. Rev., B67, 024518, 2003.

\item  M. E. McHenry, H. S. Lessure, M. P. Maley, J. Y. Coulter,
I. Tanaka, and H. Kojima, Physica C 190, 403, 1992.

\item  M. Guillot, M. Potel, P. Gougeon, H. Noel, J. C. Levet, G.
Couteau, and J.S. Tholence, Phys. Lett., A 127, 363, 1988.

\item  J. Vanacken, L. Trappeniers, K. Rosseel, I. N. Goncharov,
V.V. Moshchalkov,  Y. Bruynseraede, Physica C 332, 411, 2000.

\item G. Fuchs, C. Wenger, A. Gladun, S. Gruss, P. Schaetzle, G.
Krabbes, J. Fink, K.H. Muller and L. Schultz, EUCAS-99 Conference.

\item  Z. W. Zhao, S. L. Li, Y. M. Ni, H. P. Yang, Z. Y. Liu, H.
H. Wen, Phys. Rev. B 65,  064512, 2002.

\item K. H. Muller and C. Andrikidis, Phys. Rev. B 49, 1294, 1994.

\item  V. V. Chabanenko, A. I. D'yachenko, A. V. Chabanenko, M. V.
Zalutsky, H. Szymczak, S. Piechota and A. Nabialek, Superconduct.
Sci. Technol., 1, 1181, 1998.

\item S. Vasilev, V. V. Chabanenko, A. Nabialek, V. Rusakov, S.
Piechota and H. Szymczak, Acta Physica Polonica A, 106, 777, 2004;
S. Vasiliev, A. Nabialek, V. Chabanenko, V. Rusakov, S. Piechota,
H. Szymczak, Acta Phys. Pol. A 109, 661, 2006; A. Nabialek, S.
Vasiliev, V. Chabanenko, V. Rusakov, S. Piechota, H. Szymczak,
Acta Phys. Pol. A, 114, 2008; S. Vasiliev, A. Nabialek, V.F.
Rusakov, L.V. Belevtsov, V. V. Chabanenko and H. Szymczak,  Acta
Phys. Pol. A, 118, 2010; V. Rusakov, S. Vasiliev, V. V. Chabanenko
, A. Yurov, A. Nabialek, S. Piechota and H. Szymczak, Acta Phys.
Pol. A, 109, 2006.

\item  V. V. Chabanenko, V. F. Rusakov, V. A. Yampol'skii, S.
Piechota, A. Nabialek, S.V. Vasilev, and H. Szymczak, cond-mat.
0106379, 2002.

\item  J. I. Gittleman and B. Rosenblum, Journ. Appl. Phys., 39,
2617, 1968.

\item  R. G. Mints and A.L. Rakhmanov, J. Phys., D 12, 1929, 1979.

\item  G. Kumm, U. Ring and K. Winzer, Cryogenics, 36, 255, 1996.

\item S. Shimamoto, Cryogenics, 14, 568, 1974.

\item  I. L. Maksimov and R. G. Mints, J.Phys., D 13, 1689, 1980.

\item  N. H. Zebouni, A. Venkataram, G. N. Rao, C. G. Grenier, and
J. M. Reynolds, Phys. Rev. Lett., 13, 606, 1964.

\item  P. G. de Gennes and J. Martison, Rev. Mod. Phys., 36, 45,
1964.

\item  A. C. Bodi, I. Kirschner, R. Laiho, and E. Lahderanta,
Solid State Comm., 98, 1049, 1996.

\item  I. Legrand, I. Rosenmann, Ch. Simon, G. Collin, Physica C
211, 239, 1993.

\item  L. Legrand, I. Rosenman, R. G. Mints, G. Collin and E.
Janod, Europhys. Lett., 34, 287, 1996.

\item  A. Gurevich, Appl. Phys. Lett., 78, 1891, 2001.

\item  G. L. Dorofeev, A.B. Imenitov, and E. Yu. Klimenko,
Cryogenics 20, 307, 1980.

\item  R. G. Mints and A. L. Rakhmanov, J. Phys., D 15, 2297,
1982.

\item  D. Monier and L. Fruchter, Eur. Phys. J., B 3, 143, 1998.

\item A. M. Campbell and J. E. Evetts, Critical current in
superconductors, London, 1972.

\item A. Gurevich and H. Kupfer, Phys. Rev., B 48, 6477, 1993.

\item A. P. Malozemoff, Physica C, 185, 264, 1999.

\item Z. J. Huang, Y. Y. Xue, H. H. Feng, and C. W. Chu, Physica
C, 184, 371, 1991.

\item D. Shi and M. Xu, Phys. Rev., B 44, 4548, 1991.

\item  Y. Yeshurun and A. P. Malozemoff, Phys. Rev. Lett., 60,
2202, 1988.

\item  H. G. Schnack, R. Griessen, J. G. Lensink, C. J. van der
Beek, and P. H. Kes, Physica C 197, 337, 1992.

\item  A. Gurevich and E. H. Brandt, Phys. Rev. Lett., 73, 178,
1994.

\item  G. Blatter, M. V. Feigel'man, V. B. Geshkenbein, A. I.
Larkin, V. M. Vinokur, Rev. Mod. Phys., 66, 1125, 1994.

\item  M. P. Fisher, Phys. Rev. Lett., 62, 1415, 1989.

\item  Y. Yeshurun, A. P. Malozemoff, and A. Shaulov, Rev. Mod.
Phys., 68, 911, 1996.

\item  V. Feigel'man, V. B.Geshkenbein, and V. M. Vinokur, Phys.
Rev. Lett., 63, 2303, 1989.

\item  P. H. Kes, J. Aarto, J. van den Berg, C. J. van der Beek,
and J. A. Mydosh, Supercon. Sci.Technol., 1, 242, 1989.

\item  J. R. Tompson, Y. R. Sun, and F. Holtzberg, Phys. Rev. B44,
458, 1991.

\item  P. W. Anderson, Phys. Rev. Lett., 9, 309, 1962.

\item  P. W. Anderson and Y. B. Kim, Rev. Mod. Phys., 36, 39,
1964.

\item  C. S. Pande and R. A. Masumura, Physica C 332, 292, 2000.

\item  L. Burlachkov, D. Giller, and R. Prozorov, condmat-9802076,
1998.

\item  E. Zeldov, N. M. Amer, G. Koren, A. Gupta, M. W. McElfresh
and R.J. Gambino, Appl. Phys. lett., 56, 680, 1990.

\item  E. H. Brandt, Phys. Rev. B 55, 14513, 1997.

\item  Y. Mawatari, A. Saws, H. Obara, M. Umeda, and H. Yamasaki,
Appl. Phys. Lett., 70, 2300, 1997.

\item  A. Gerber, J. N. Li, Z. Tarnawski, J. J. M. Franse, and A.
A. Menovsky, Phys. Rev., B 47, 6047, 1993.

\item  Z. Wang and D. Shi, Solid State Commun., 90, 405, 1994.

\item  T. Nattermann, Phys. Rev. Lett., 64, 2454, 1990.

\item N. A. Taylanov, cond-mat., 0207009, 2002.

\item  R. P. Huebener, Super. Sci. Technol., 8, 189, 1995.

\item  T. Sasaki, K. Watanabe, and N. Kobayashi, Sci. Rep., RITU A
42, 351, 1996.

\item A. T. Fiory and B. Serin, Phys. Rev. Lett., 16, 308, 1966.

\item  A. T. Dorsey, Phys. Rev. B 46, 8376, 1992.

\item  P. Ao, J. Low. Temp. Phys., 107, 347, 1997.

\item  R. P. Huebener, and A. Seher, Phys. Rev., 181, 701, 1969.

\item  S. N. Artemenko, and A. F. Volkov, JETP, 70, 1051, 1976.

\item  A. V. Gurevich and R.G. Mints, JETP Lett., 31, 52, 1980.

\item  A. N. Larkin and Yu. N. Ovchinnikov, JETP, 49, 2337, 1981.

\item  V. L. Ginzburg and G. F. Jarkov, Uspekh. Fiz. Nauk, 125,
19, 1978.

\item  N. B. Kopnin, Zh. Eksp. Teor. Fiz., 69, 364, 1975.

\item  R. Solomon and F. A. Otter, Phys. Rev., 164, 608, 1967.

\item  C. Caroli, P. G. DeGennes, and J. Matrison, Phys. Lett., 9,
307, 1964.

\item  L. D. Landau and E. M. Lifshitz, Statistical Physics,
Moscow, Nauka, 1976.

\item  C. A. Duran, P. L. Gammel, R. E. Miller, D. J. Bishop,
Phys. Rev., B 52, 75, 1995.

\item  V. Vlasko-Vlasov, U. Welp, V. Metlushko, and G.W. Crabtee,
Physica C, 341-348, 1281, 2000.

\item  T. H. Johansen, M. Basiljevich, D. V. Shantsev, P. E. Goa,
Y. M. Galperin, W. N. Kang, H. J. Kim, E. M. Choi, M. S. Kim, S.I.
Lee, Europhys. Lett., 59, 599, 2002.

\item  A. V. Bobyl, D. V. Shantsev, Y. M. Galperin, A. A. F.
Olsen, T. H. Johansen, W. N. Kang and S. I. Lee, cond-mat.
0304603, 2003.

\item  F. L. Barkov, D. V. Shantsev, T. H. Johansen, W. N. Kang,
H. J. Kim, E. M. Choi, and S. I. Lee, Phys. Rev., B 67, 064513,
2003.

\item  P. Leiderer, J. Boneberg, P. Brull, V. Bujok, S.
Herminghaus, Phys. Rev. Lett., 71, 2646, 1993.

\item  U. Bolz, D. Schmidt, B. Biehlera, B.U. Runge, R.G. Mints,
K. Numssen, H. Kinder, P. Leiderer, Physica C 388, 715, 2003.

\item  U. Bolz et al., Europhys. Lett., 64, 517, 2003.

\item  M. S. Welling et al., Physica C 411, 11, 2004.

\item  I. A. Rudnev et al., Cryogenics, 43, 663, 2003.

\item A. L. Rakhmanov, D. V. Shantsev, Y. M. Galperin and T.H.
Johansen, cond-mat/0405446, 2004.

\item  I. Aronson, A. Gurevich, and V. V. Vinokur, Phys. Rev.
Lett., 87, 067003, 2001.

\item  I. S. Aranson et al., Phys. Rev. Lett. 94, 037002, 2005.

\item D. V. Denisov, A. L. Rakhmanov, D. V. Shantsev, Y. M.
Galperin, T. H. Johansen, Phys. Rev. B 73, 014512 2006.

\item M. R. Wertheimer and J. G. Gilchrist, J. Phys. Chem. Solids
28, 2509, 1967.

\item  G. T. Seidler, C. S. Carrillo, T. F. Rosenbaum, U. Welp, G.
W. Crabtree, and V. M. Vinokur,  Phys. Rev. Lett., 70, 2814, 1993.

\item  P. Bak, C. Tang and K. Wiesenfeld, Phys. Rev. Lett., 59,
381, 1987.

\item  P. Bak, C. Tang and K. Wiesenfeld, Phys. Rev., A 38, 3645,
1988.

\item  R. J. Zieve, T. F. Rosenbaum, H. M. Jaeger, G. T. Seidler,
G. W. Crabtree, and U. Welp, Phys. Rev., B 53, 11849, 1996.

\item C. J. Olson, C. Reichhardt and F. Nori, Phys. Rev., B 56,
6175, 1998.

\item  K. E. Bassler and M. Paczuski, Phys. Rev. Lett., 81, 3761,
1998.

\item  S. Field, J. Witt and F. Nori, Phys. Rev. Lett., 74, 1206,
1995.

\item  O. Pla and F. Nori, Phys. Rev. Lett., 67, 919, 1991.

\item  R. A. Richardson, O. Pla and F. Nori, Phys. Rev. Lett., 72,
1268, 1994.

\item  R. Cruz, R. Mulet and E. Altshuler, Physica A 275, 15,
2000.

\item  G. Mohler and D. Stroud, Phys. Rev., B 60, 9738, 1999.

\item  R. Mulet and E. Altshuler, Physica C 281, 317, 1997.

\item  C. M. Aegerter, M. S. Welling, and R. J. Wijngaarden,
cond-mat. 0305591, 2003.

\item  W. Barford, Phys. Rev., B 56, 425, 2002.

\item  E. Altshuler and T. H. Johansen, cond-mat. 0402097, 2004.

\item  K. Behina, C. Capan, D. Mailly, and B. Etienne, Phys. Rev.,
B 61, 3815, 2000.

\item  S. Majumdar, M. R. Lees, G. Balakrishnan, and D. McK Paul,
cond-mat. 0305208, 2003.

\item  A. Milner, Physica B 294, 388, 2001.

\item  A. Gerber, A. Milner, Phys. Rev., B 62, 9753, 2000.

\item  A. Terentiev, D. B. Watkins, L. E. De Long, L. D. Cooley,
D. J. Morgan, and J. B. Ketterson, Phys. Rev., B 61, R9249, 2000.

\item S. B. Roy and P. Chaddah, Physica C 279, 1, 1997.

\item  H. Safar, P. L. Gammel, D. A. Huse, D. J. Bishop, W. C.
Lee, and D. M. Ginsberg, Phys. Rev. Lett. 70, 3800, 1993.

\item W. K. Kwok, J. A. Fendrich, C. J. van der Beek, and G. W.
Crabtree, Phys. Rev. Lett. 73, 2614, 1994.

\item P. L. Gammel, U. Yaron, Y. P. Ramirez, D.J . Bishop, A. M.
Chang, R. Ruel, L. N. Pfeiffer, E. Bucher, G. D’Anna, D. A. Huse,
K. Mortensen, M. R. Eskildsen, and P. H. Kes, Phys. Rev. Lett. 80,
833, 1998.

\item M. Daemling, J. M. Seutjens, and D. C. Laralestier, Nature
(London) 346, 332, 1990.

\item L. Klein, E. R. Yacoby, Y. Yeshurun, A. Erb, G. Muller-Vogt,
V. Breit, and H. Wuhl, Phys. Rev. B 49, 4403, 1994.

\item L. Krusin-Elbaum, L. Civale, V. M. Vinokur, and F.
Holtzberg, Phys. Rev. Lett. 69, 2280, 1992.

\item Y. Abulafia, A. Shaulov, Y. Wolfus, R. Prozorov, L.
Burlachkov, Y. Yeshurun, D. Majer, E. Zeldov, H. Wuhl, V.B.
Geskenbein, and V. M. Vinokur, Phys. Rev. Lett. 77, 1596, 1996.

\item D. Giller, A. Shaulov, Y. Yeshurun, and J. Giapintzakis,
Phys. Rev. B 60, 106, 1999.

\item D. Giller, A. Shaulov, R. Prozorov, Y. Abulafia, Y. Wolfus,
L. Burlachkov, Y. Yeshurun, E. Zeldov, and V. M. Vinokur, J. L.
Peng, and R. L. Greene, Phys. Rev. Lett. 79, 2542, 1997.

\item Y. V. Bugoslavsky, A. L. Ivanov, A. A. Minakov, and S.I.
Vasyurin, Physica C 233, 67, 1994.

\item V. Hardy, A. Wahl, A. Ruyter, A. Maignan, C. Martin, L.
Coudrier, J. Provost, and C. Simon, Physica C 232, 347, 1994.

\item A. Maignan, S.N. Putilin, V. Hardy, C. Simon, and B. Raveau,
Physica C 266, 173, 1996.

\item T. Aouaroun and C. Simon, Physica C 306, 238, 1998.

\item M. Pissas, D. Stampoulos, E. Moraitakis, G. Kallias, D.
Niarchos, and M. Charalambous, Phys. Rev. B 59, 121, 1999.

\item V. Yeshurun, N. Bontemps, L. Burlachkov and A. Kapitulnik,
Phys. Rev. B 49, 1548, 1994.

\item R. Yoshizaki, I. Ikeda  and J. D. Jeon, Physica C 225, 299,
1994.

\item G. Yang, P. Shang, S. D. Sutton, I. P. Jones, J. S. Abell,
and C. E. Gough, Phys. Rev. B 48, 4054, 1993.

\item F. Zuo, S. Khizroev, G. C. Alexandrakis, and V. N. Kopylov,
Phys. Rev. B 52, R755, 1995.

\item N. Chikumoto, M. Konczykowski, N. Motohira, and A. P.
Malozemoff, Phys. Rev. Lett. 69, 1260, 1992.

\item Y. Kopelevich, and P. Esquinazi, J. Low Temp. Phys., 113,
1998.

\item  P. Esquinazi, A. Setzer, D. Fuchs, Y. Kopelevich, E. Zeldov
and C. Assmann, Phys. Rev., B 60, 17, 12454, 1999.

\item  A. Yu. Galkin, Y. Kopelevich, P. Esquinazi, A. Setzer, V.M.
Pan, and S. N. Barilo, cond-mat. 9912346, 1999.

\item A. Pan, M. Ziese, R. Hohne, P. Esquinazi, S. Knappe, and H.
Koch, Physica C 301,72 1998; A. Pan, R. Hohne, M. Ziese, P.
Esquinazi, and C. Assmann, in Physics and Material Sciences of
Vortex States, Flux Pinning and Dynamics, edited by R. Kossowsky
et al.(Kluwer Academic), Dordrecht, 545, 1999.

\item  Y. Kopelevich, S. Moehlecke, J. H. Torres, R. Ricardo da
Silva, and P. Esquinazi, J. Low. Temp. Phys., 116, 261, 1999.

\item D. Stamopoulos, A. Speliotis, and D. Niarchos, Supercond.
Sci. Technol., 17, 1261, 2004.

\item  A. I. Larkin and Yu. N. Ovchinnikov, J. Low Temp. Phys. 34,
409, 1979.

\item A. A. Zhukov, H. Kupfer, G. Perkins, L. F. Cohen, A. D.
Caplin, S. A. Klestov, H. Claus, V. I. Voronkova, T. Wolf and H.
Wuhl, Phys. Rev., B 51, 12704, 1995.

\item  B. Khaykovich, E. Zeldov, D. Majer, T. W. Li, P. H. Kes,
and M. Konczykowski, Phys. Rev. Lett. 76, 2555 1996; E. Zeldov, A.
Larkin, V. Geshkenbein, M. Konczykowski, D. Majer, B. Kheykovich,
V. Vinokur, and H. Shtrikham, Phys. Rev. Lett. 73, 1428, 1994.

\item  A. Schilling, R. A. Fisher, N. E. Phillips, U. Welp, D.
Dasgupta, W. K. Kwok, and G. W. Crabtree, Nature 382, 791 1996.

\item  R. Cubitt, E. M. Forgan, G. Yang, S. L. Lee, D. M. Paul, H.
A. Mook, M. Yethiraj, P. H. Kes, T. W. Li, A. A. Menovsky, Z.
Tarnawski, and K. Mortensen, Nature (London) 365, 407 1993.

\item  S. L. Lee et al., Phys. Rev. Lett. 71, 3862, 1993.

\item  D. Ertas and D. R. Nelson, Physica (Amsterdam) 272C, 79,
1996.

\item  V. Vinokur, B. Khaykovich, E. Zeldov, M. Konczykowski, R.A.
Doyle, and P. Kes, Physica C 295, 209, 1998.

\item  T. Giamarchi and P. Le Doussal, Phys. Rev. B 55, 6577,
1997.

\item T. Giamarchi and P. Le Doussal, Phys. Rev. Lett. 72, 1530,
1994.

\item J. Kierfeld, T. Nattermann, and T. Hwa, Phys. Rev. B 55,
626, 1997.

 \item D. S. Fisher, Phys. Rev. Lett. 78, 1964, 1997.

\item S. Kokkaliaris, P. A. J. de Groot, S. N. Gordeev, A. A.
Zhukov, R. Gagnon, and L. Taillefer, Phys. Rev. Lett. 82, 5116,
1999.

\item  Y. Paltiel, E. Zeldov, Y. Myasoedov, M. L. Rappaport, G.
Jung, S. Bhattacharya, M. J. Higgins, Z. L. Xiao, E. Y. Andrei, P.
L. Gammel, and D.J. Bishop, Phys. Rev. Lett. 85, 3712, 2000.

\item  T. Halpin-Healy, Y. C. Zhang, Phys. Rep. 254, 215, 1995.

\item  T. Nishizaki, T. Naito, and N. Kobayashi, Phys. Rev. B 58,
11169, 1998.

\item  T. Nishizaki, T. Naito, S. Okayasu, A. Iwase, and N.
Kobayashi, Phys. Rev. B 61, 3649, 2000.

\item  J. Karpinski, S. Kazakov, M. Angst, A. Mironov, M. Mali,
and J. Roos, Phys. Rev. B 64, 094518, 2001.}
\end{enumerate}

}\end{multicols}

\end{document}